\documentclass[12pt]{scrartcl}

\usepackage[centertags]{amsmath}
\usepackage{amssymb}
\usepackage{hyperref}
\usepackage{cite}
\hypersetup{
    bookmarks=false,      
    pdfstartview={FitH},  
    colorlinks=true,      
    linkcolor=black,      
    citecolor=blue,       
    filecolor=blue,       
    urlcolor=blue         
}

\usepackage{graphicx}
\usepackage{varioref}
\usepackage{paralist}
\usepackage{subfigure}
\usepackage{lipsum}
\usepackage{scrtime}
\usepackage[table]{xcolor}
\usepackage{pifont}

\labelformat{section}{Section #1} 
\labelformat{subsection}{Section #1} 
\labelformat{equation}{Eq.~(#1)} 
\labelformat{figure}{Fig.~#1} 
\labelformat{subfigure}{Fig.~\thefigure#1} 
\labelformat{table}{Tab.~#1} 

\setkomafont{subsection}{\normalcolor\bfseries}
\addtokomafont{caption}{\footnotesize}

\newcommand{\vev}{{\small{\textit{vev}}}}
\newcommand{\vevof}[1]{\ensuremath{\langle #1 \rangle}}
\newcommand{\vevs}{{\small{\textit{vevs}}}}
\newcommand{\SU}[1]{\ensuremath{\mathrm{SU}(#1)}}
\newcommand{\U}[1]{\ensuremath{\mathrm{U}(#1)}}
\newcommand{\bs}[1]{\ensuremath{\boldsymbol{#1}}}

\newcommand{\UPMNS}{\ensuremath{U_\mathrm{PMNS}}}

\newcommand{\Cone}{\ensuremath{y_{2}}}
\newcommand{\Ctwo}{\ensuremath{y_{1}}}
\newcommand{\Cthree}{\ensuremath{C^\nu_{1}}}
\newcommand{\Cfour}{\ensuremath{C^\nu_{2}}}
\newcommand{\Cfive}{\ensuremath{C^\nu_{5}}}
\newcommand{\Csix}{\ensuremath{C^\nu_{6}}} 
\newcommand{\Cseven}{\ensuremath{C^\nu_{3}}}
\newcommand{\Ceight}{\ensuremath{C^\nu_{4}}}
\newcommand{\Cnine}{\ensuremath{y_{e}}}
\newcommand{\Cten}{\ensuremath{C^e_{1}}}
\newcommand{\Celeven}{\ensuremath{C^e_{2}}}
\newcommand{\Ctwelve}{\ensuremath{C^e_{3}}}

\newcommand{\Cfourteen}{\ensuremath{y_{\mu}}}
\newcommand{\Cfifteen}{\ensuremath{C^\mu_{1}}}
\newcommand{\Csixteen}{\ensuremath{C^\mu_{2}}}
\newcommand{\Cseventeen}{\ensuremath{C^\mu_{3}}}

\newcommand{\Cnineteen}{\ensuremath{y_{\tau}}}
\newcommand{\Ctwenty}{\ensuremath{C^\tau_{1}}}
\newcommand{\Ctwentyone}{\ensuremath{C^\tau_{2}}}
\newcommand{\Ctwentytwo}{\ensuremath{C^\tau_{3}}}

\newcommand{\hu}{\ensuremath{h_{u}}}
\newcommand{\hd}{\ensuremath{h_{d}}}

\newcommand{\ec}{\ensuremath{e^{c}}}
\newcommand{\muc}{\ensuremath{\mu^{c}}}
\newcommand{\tauc}{\ensuremath{\tau^{c}}}

\newcommand{\phit}{\ensuremath{\widetilde{\varphi}}}

\newcommand{\be}{\begin{equation}}
\newcommand{\ee}{\end{equation}}
\newcommand{\ol}{\overline}
\def\bea{\begin{eqnarray}}
\def\eea{\end{eqnarray}}
\newcommand{\wt}{\widetilde}

\begin{document}

\pagenumbering{roman}

\titlehead{
\begin{flushright}
LPSC-12250\\
IPPP-12-61\\
DCPT-12-122
\end{flushright}
}

\title{\Large A Minimal Model of Neutrino Flavor}
\author{Christoph Luhn\footnote{christoph.luhn@durham.ac.uk}\\
{\normalsize \it Institute for Particle Physics Phenomenology}\\[-1ex]
{\normalsize \it University of Durham, Durham DH1 3LE, UK}\\
\and
Krishna Mohan Parattu\footnote{krishna@iucaa.ernet.in}\\
{\normalsize \it Inter-University Centre for Astronomy and Astrophysics}\\[-1ex]
{\normalsize \it Ganeshkhind, Pune 411007, India}\\
\and
Ak\i{}n Wingerter\footnote{akin@lpsc.in2p3.fr}\\
{\normalsize \it Laboratoire de Physique Subatomique et de Cosmologie}\\[-1ex]
{\normalsize \it UJF Grenoble 1, CNRS/IN2P3, INPG}\\[-1ex]
{\normalsize \it 53 Avenue des Martyrs, F-38026 Grenoble, France}\\
}
\date{}
\dedication{\vspace{-2ex}{\bfseries Abstract}\\[2ex]
\begin{minipage}[b]{0.9\linewidth} 
\small 
Models of neutrino mass which attempt to describe the observed lepton
mixing pattern are typically based on discrete family symmetries with
a non-Abelian and one or more Abelian factors. The latter so-called
shaping symmetries are imposed in order to yield a realistic
phenomenology by forbidding unwanted operators. Here we propose a
supersymmetric model of neutrino flavor which is based on the
group~$T_7$ and does not require extra $\mathbb Z_N$ or $\U{1}$
factors, which makes it the smallest realistic family symmetry that
has been considered so far.  At leading order, the model predicts
tribimaximal mixing which arises completely accidentally from a
combination of the $T_7$ Clebsch-Gordan coefficients and suitable
flavon alignments.  Next-to-leading order (NLO) operators break the
simple tribimaximal structure and render the model compatible with the
recent results of the Daya Bay and Reno collaborations which have
measured a reactor angle of around $9^\circ$. Problematic NLO
deviations of the other two mixing angles can be controlled in an
ultraviolet completion of the model.
\end{minipage} }

\maketitle[0]
\thispagestyle{empty}
\newpage

\pagenumbering{arabic}

\clearpage
\newpage
\section{Introduction}

The triplication of chiral families remains one of the biggest mysteries in
particle physics. A clue towards unraveling the principle behind
the fact that quarks and leptons come in three copies is provided by the
observation of a remarkable lepton mixing pattern: contrary to the quark
sector, the mixing of the leptons is described by two large and one small
angle. Until recently, the results of neutrino oscillation experiments were
well compatible with a PMNS matrix of the intriguingly simple
structure~\cite{Harrison:2002er,Harrison:2002kp,Xing:2002sw}
\be
U_{\mathrm{PMNS}} ~\approx ~ \begin{pmatrix}
\frac{2}{\sqrt{6}} &\frac{1}{\sqrt{3}}&0\\
-\frac{1}{\sqrt{6}} &\frac{1}{\sqrt{3}}&\frac{1}{\sqrt{2}}\\
-\frac{1}{\sqrt{6}} &\frac{1}{\sqrt{3}}&-\frac{1}{\sqrt{2}}
\end{pmatrix}.
\ee
This so-called tribimaximal mixing suggests an intimate connection of the
three generations of leptons, which can be realized in the framework of
non-Abelian discrete family symmetries. Imposing such a horizontal symmetry~$G$
allows to unify different generations into a multiplet of the given
non-Abelian group. With three families, the physically interesting groups
should have a triplet representation, limiting possible choices to subgroups of
$U(3)$. Many of these have been successfully applied to
construct models of tribimaximal lepton mixing, see for 
instance~\cite{Altarelli:2010gt,Ishimori:2010au,Grimus:2011fk} and
references therein.

In general, family symmetry models can be classified according to the origin
of the symmetry of the neutrino mass matrix. We will assume that neutrinos are
Majorana particles; then their mass matrix is always symmetric under a Klein
symmetry $\mathbb Z_2 \times \mathbb Z_2$. Working in a basis where the
charged leptons are diagonal,\footnote{This classification can also be applied
to GUT models where the charged lepton mass matrix is typically only 
approximately diagonal. In such a setup, the total PMNS mixing
matrix will involve charged lepton corrections that have to be
taken into account separately, see
e.g.~\cite{Hagedorn:2010th,Cooper:2010ik,Cooper:2012wf,Hagedorn:2012ut,King:2012in}, 
leading to characteristic mixing sum rules~\cite{King:2005bj,King:2007pr}.}
the explicit form of the Klein symmetry expressed in terms of $3\times 3$
matrices is dictated by the PMNS mixing matrix and can be determined as
\be
U^\ast_{\mathrm{PMNS}} 
\begin{pmatrix}
(-1)^p &0&0\\
0&(-1)^q&0\\
0&0&(-1)^{p+q}
\end{pmatrix}
U^T_{\mathrm{PMNS}}
\ .\label{eq:klein}
\ee
Here $p=0,1$ and $q=0,1$, yielding a symmetry group of four elements.
This neutrino flavor symmetry can arise as a residual symmetry of the
underlying family symmetry~$G$, in other words, the four elements
of~\ref{eq:klein} can also be elements of the imposed family symmetry. Models of
this type are called  direct models~\cite{King:2009ap}. In indirect models, on
the other hand, the above Klein symmetry is not a subgroup of~$G$. Models of
this class are typically based on the type~I seesaw
mechanism~\cite{Minkowski:1977sc,Gell-Mann,Yanagida,Mohapatra:1979ia} with the
assumption of sequential dominance~\cite{King:1998jw,King:1999cm,King:2002nf},
i.e. a hierarchy among the three terms arising from three right-handed
neutrinos. Here, the main role of the family symmetry consists in explaining
special vacuum configurations of the flavon
fields that break the family symmetry~\cite{King:2009ap}. In addition to these pure classes of models there
are semi-direct models in which one of the $\mathbb Z_2$ factors of the Klein symmetry
arises as a residual symmetry of~$G$, while the other factor arises accidentally. In fact,
the famous Altarelli-Feruglio model based on $A_4$~\cite{Altarelli:2005yp,Altarelli:2005yx}
belongs to the semi-direct class as the tribimaximal $\mu$-$\tau$ symmetry
is not part of $A_4$. 

Hitherto, regardless of the type of model, the non-Abelian family symmetry has always been
augmented by extra Abelian factors such as $\mathbb Z_N$ and $\U{1}$ in order 
to yield realistic phenomenology. These shaping symmetries were crucial for
controlling the coupling of the Standard Model neutral flavon fields to
the leptons. In~\cite{Altarelli:2005yp,Altarelli:2005yx}, for instance,  the
neutrino and charged lepton sectors are separated by means of a $\mathbb Z_3$
shaping symmetry. In that sense, one should speak more precisely of the
Altarelli-Feruglio $A_4 \times \mathbb Z_3$  model, and the family symmetry $G$ is
thus a group of order 36, rather than 12. 

Mindful of this subtlety of defining the full family symmetry of a model, a
systematic scan over 76 different groups has been performed
in~\cite{Parattu:2010cy} with the purpose of studying whether or not
there is an inherent connection between $A_4$ and tribimaximal mixing. The
results of the scan proved that there are indeed several, even more minimal,
groups which are capable of describing tribimaximal mixing (assuming the
imposed simple alignment of the flavon \vevs{} can be justified). The smallest
such group was identified to be $T_7$, a group of order 21 which is sometimes
also called the Frobenius group $\mathbb Z_7 \rtimes \mathbb
Z_3$~\cite{Luhn:2007sy,Luhn:2007yr,Hagedorn:2008bc}. Note that this non-Abelian symmetry
does not require any extra shaping symmetry. It is furthermore interesting to
point out that $T_7$ has no $\mathbb Z_2$ subgroups, and hence 
the Klein symmetry of the neutrino mass matrix cannot be part
of this family symmetry. As such, a
corresponding $T_7$ model would be neither direct nor semi-direct, but rather
of indirect type. Not necessarily requiring the seesaw mechanism, it would --
to the best of our knowledge --  be the first indirect model which is not based on
the assumption of sequential dominance and the quadratic appearance of flavon fields in the
effective neutrino mass term. 

It is the purpose of this article to present the details of a complete $T_7$
model of leptons which 
yields tribimaximal mixing at leading order, including a discussion of the
vacuum alignment. We put particular emphasis on
explaining how the Klein symmetry arises completely accidentally from a
combination of the $T_7$ Clebsch-Gordan (CG) coefficients and suitable flavon
alignments. As tribimaximal mixing has been ruled out by the Daya
Bay~\cite{An:2012eh} and Reno~\cite{Ahn:2012nd} measurements of a reactor
mixing angle $\theta_{13}$ of around $9^\circ$, it is necessary to investigate if 
next-to-leading order (NLO) effects can generate large enough deviations from the
tribimaximal leading order prediction. With the Klein symmetry
arising purely accidentally, NLO terms are bound to perturb the tribimaximal
mixing pattern. While switching on $\theta_{13}$, NLO corrections will in
general also give rise to perturbations of the atmospheric and, more
critically, the solar  mixing angle which may be phenomenologically
unacceptable. However, these unwanted NLO corrections can be suppressed in an
ultraviolet completed version of the $T_7$ model involving special messenger
fields. 

The remainder of this article is organized as follows. We define and discuss the minimal
$T_7$ model in~\ref{sec:TBMLO}, both at leading as well as next-to-leading
order. The ultraviolet completion of the model is presented
in~\ref{sec:uv-mo}. \ref{sec:vev-4} addresses the question of vacuum
alignment, and the conclusions are drawn in~\ref{sec:concl}. The relevant
group theoretic details of $T_7$ are laid out in~\ref{app:A}, including the
generators of the  representations in a basis with a diagonal order three
element as well as the corresponding CG coefficients.

\section{The minimal $\bs{T_7}$ model}
\label{sec:TBMLO}

\subsection{Tribimaximal mixing at leading order}

In this section we describe a minimal model of leptons based on the family
symmetry~$T_7$. This non-Abelian finite 
group comprises 21 elements and has three singlet as well as two triplet
representations, which we denote by ${\bf 1}$, ${\bf 1'}$, ${\bf 1''}$, ${\bf 3}$
and~${\bf \ol 3}$, respectively.   
As the model does not feature any shaping symmetries, its structure is
solely determined by the family symmetry $T_7$ as well as the gauge
symmetry $\SU{2}_L\times \U{1}_Y$. We work in a supersymmetric framework with
two Higgs doublets, $\hu{}$ and $\hd{}$, transforming trivially under $T_7$. The
three generations of left-handed lepton doublets $L$ are unified 
in a triplet representation of $T_7$, while the right-handed charged
leptons $\ec{}$, $\muc{}$ and $\tauc{}$ live in the three distinct one-dimensional
representations. In order to break the family symmetry, two flavon
fields are introduced, namely $\wt \varphi$ and $\varphi$ transforming as a ${\bf \ol 3}$ and 
${\bf 3}$, respectively. The particle content of this model
 is summarized in \vref{tab:minmodel}. For later purposes we have also listed
the hypercharges as well as the charges under the $\U{1}_R$ symmetry
which is required for the $F$-term flavon alignment mechanism discussed in \ref{sec:vev-4}.
\begin{table}[t]
\begin{center}
\begin{tabular}{|c||c|c|c|c|c|c||c|c|}\hline
$\phantom{\Big|}$Field$\phantom{\Big|}$ & $L$&$\ec{}$&$\muc{}$&$\tauc{}$&$\hu{}$&$\hd{}$ &$\wt \varphi$&$\varphi$ \\\hline
$\phantom{\Big|}$$T_7$$\phantom{\Big|}$ & ${\bf 3}$ & ${\bf 1}$ & ${\bf 1'}$ & ${\bf 1''}$ & ${\bf 1}$ & ${\bf
  1}$ & ${\bf  \ol 3}$& ${\bf 3}$  \\\hline
$\phantom{\Big|}$$\U{1}_Y$ $\phantom{\Big|}$& $-1$ & $2$ & $2$ & $2$ & $1$ & $-1$ & $0$ & $0$ 
 \\\hline
$\phantom{\Big|}$$\U{1}_R$$\phantom{\Big|}$ & $1$ & $1$ & $1$ & $1$ & $0$ & $0$ & $0$ & $0$ \\\hline
\end{tabular}
\end{center}
\caption{\label{tab:minmodel}The transformation properties of the minimal
  $T_7$ model of leptons.}
\end{table}

With these $T_7$ assignments, the leading order superpotentials of the
charged lepton and the neutrino sector read
\bea
W_\ell &=& 
\Cnine\mbox{$\frac{\wt \varphi}{\Lambda}$} L\ec{}\hd{} + 
\Cfourteen\mbox{$\frac{\wt \varphi}{\Lambda}$} L\muc{}\hd{} +
\Cnineteen\mbox{$\frac{\wt \varphi}{\Lambda}$} L\tauc{}\hd{} \ ,
\label{eq:leading-order-superpotential-ell}\\[2mm]
W_\nu &=&\Ctwo \mbox{$\frac{\wt \varphi}{\Lambda^2}$}  LL \hu{}\hu{} +
\Cone\mbox{$ \frac{\varphi}{\Lambda^2}$}  LL \hu{}\hu{} \ ,
\label{eq:leading-order-superpotential-nu}
\eea
respectively. Here, the $y$'s are dimensionless coupling constants,
and $\Lambda$ denotes a common cut-off above the scale of family
symmetry breaking. The neutrino mass terms originate from the Weinberg
operator~\cite{Weinberg:1979sa} augmented by a flavon field.

It is important to note that only one flavon, $\wt\varphi$, enters the charged
lepton sector at leading order. The other flavon, $\varphi$, cannot couple to
the charged leptons due to the absence of a singlet representation in the
$T_7$ Kronecker product ${\bf 3\otimes 3}$, see \ref{app:A}. We will see below that the \vev{} of
$\wt\varphi$ can be chosen such that it breaks $T_7$ down to $\mathbb Z_3$,
thus leading to a diagonal charged lepton mass matrix.
In the neutrino sector, both flavons are present, as 
the symmetric product of ${\bf 3\otimes 3}$  contains both a ${\bf 3}$ and a
${\bf \ol 3}$. 

In order to find the vacuum alignments that lead to tribimaximal
mixing, we insert the flavon \vevs{} in
\ref{eq:leading-order-superpotential-ell} and
\ref{eq:leading-order-superpotential-nu} in their most general
form. Contracting the $T_7$ indices using the CG coefficients given
in \ref{app:A} and inserting the Higgs \vevs{} $v_u$ and $v_d$, we
obtain the respective mass matrices. In the left-right convention for
the charged leptons, where the left-handed particles are to the left
of the Yukawa matrix, we find
\bea
M_\ell &\!\!=\!\!& 
\left[\!\begin{pmatrix}\Cnine&0&0\\0&\Cfourteen&0\\0&0&\Cnineteen \end{pmatrix} 
\!\mbox{$\frac{\langle\wt \varphi_1 \rangle}{\Lambda}$} \,+\,
\begin{pmatrix}0&0&\Cnineteen\\\Cnine&0&0\\0&\Cfourteen&0\end{pmatrix} 
\!\mbox{$\frac{\langle\wt \varphi_2 \rangle}{\Lambda}$} \,+\,
 \begin{pmatrix}0&\Cfourteen&0\\0&0&\Cnineteen\\\Cnine&0&0\end{pmatrix}
\!\mbox{$\frac{\langle\wt \varphi_3 \rangle}{\Lambda}$}
\!\right]
\mbox{$\frac{v_d}{\sqrt{3}}$}  \ ,
\label{eq:mellLO}\\[5mm]
M_\nu &\!\!=\!\!& 
\Ctwo 
\left[\!\begin{pmatrix}1&0&0\\0&0&1\\0&1&0\end{pmatrix}
 \!\mbox{$\frac{\langle \wt \varphi_1 \rangle}{\Lambda}$} \,+\,
\begin{pmatrix}0&\omega&0\\\omega&0&0\\0&0&\omega\end{pmatrix} 
\!\mbox{$\frac{\langle \wt \varphi_2 \rangle}{\Lambda}$} \,+\,
\begin{pmatrix}0&0&\omega^2\\0&\omega^2&0\\\omega^2&0&0\end{pmatrix}
\!\mbox{$\frac{\langle \wt \varphi_3 \rangle}{\Lambda}$}\! \right]
\mbox{$\frac{v_u^2}{3\Lambda}$}  \label{eq:mnuLO}\\[2mm]
&& \hspace{-4.6mm}+ ~ \Cone
\left[\!\begin{pmatrix}2&0&0\\0&0&-1\\0&-1&0\end{pmatrix}
 \!\mbox{$\frac{\langle \varphi_1 \rangle}{\Lambda}$} \,+\,
\begin{pmatrix}0&0&-1\\0&2&0\\-1&0&0\end{pmatrix} 
\!\mbox{$\frac{\langle \varphi_2 \rangle}{\Lambda}$} \,+\,
\begin{pmatrix}0&-1&0\\-1&0&0\\0&0&2\end{pmatrix}
\!\mbox{$\frac{\langle \varphi_3 \rangle}{\Lambda}$}\! \right]
\mbox{$\frac{v_u^2}{3\sqrt{2}\Lambda}$} \, .~~~~\notag
\eea
This simple pattern suggests the following flavon vacuum configuration to
obtain tribimaximal mixing:
\be
\langle \wt \varphi \rangle ~=~ v_{\wt \varphi} \begin{pmatrix}1\\0\\0 \end{pmatrix}\ , \qquad
\langle \varphi \rangle ~=~ v_\varphi \: \mbox{$\frac{1}{\sqrt{3}}$} \begin{pmatrix}1\\1\\1 \end{pmatrix}.
\label{eq:alignments}
\ee
The alignment of $\wt \varphi$ ensures a diagonal charged lepton mass
matrix in \ref{eq:mellLO}, and brings the first line in
\ref{eq:mnuLO} into tribimaximal form.
The alignment of $\varphi$ is  of an extremely simple form
and generates a contribution to $M_\nu$ which, again, has
tribimaximal structure.
Adopting the flavon alignments of \ref{eq:alignments}, our minimal
$T_7$ model therefore predicts the following charged lepton and
neutrino mass matrices at leading order,
\bea
M_\ell &=& 
\begin{pmatrix}\Cnine&0&0\\0&\Cfourteen&0\\0&0&\Cnineteen \end{pmatrix} 
 \mbox{$ \frac{v_{\wt \varphi}}{\Lambda} \: \frac{v_d}{\sqrt{3}}$} \ ,\\[5mm]
M_\nu &=& \left[
\Ctwo \begin{pmatrix}1&0&0\\0&0&1\\0&1&0\end{pmatrix}
 \mbox{$\frac{v_{\wt \varphi}}{\Lambda}$}
~+ ~ \Cone \begin{pmatrix}2&-1&-1\\-1&2&-1\\-1&-1&2\end{pmatrix}
\mbox{$\frac{v_\varphi}{\sqrt{6}\Lambda} $}
\right] 
\mbox{$\frac{v_u^2}{3 \Lambda}$} \ ,
\label{eq:LOneutrinomassmatrix}
\eea
implying a tribimaximal PMNS mixing matrix and the complex valued neutrino
masses
\be
(m^\nu_1,m^\nu_2,m^\nu_3) ~=~
(
2 y_1 v_{\wt\varphi} + \sqrt{6} y_2 v_{\varphi} \,,\, 
2 y_1 v_{\wt\varphi} \,,\,
-2 y_1 v_{\wt\varphi} + \sqrt{6} y_2 v_{\varphi} ) 
\mbox{$\frac{v_u^2}{6\Lambda^2}$}\ .
\ee 

Before continuing to the study of NLO effects, let us pause for a
moment and appreciate the beauty of the leading order model. With
regard to the charged lepton sector, we observe that the alignment of
${\wt\varphi}$ is left invariant by the diagonal order-three
generator $d$~of~$T_7$, see \ref{tab:basis} in \ref{app:A}. The
corresponding $\mathbb Z_3$ symmetry which is thus preserved in the
charged lepton sector after the family symmetry is broken enforces a
diagonal $M_\ell$. In contrast, the tribimaximal Klein symmetry
$\mathbb Z_2\times \mathbb Z_2$ of $M_\nu$ arises completely
accidentally from a combination of the $T_7$ CG coefficients and
suitable flavon alignments (which will be derived and thus justified
in \ref{sec:vev-4}). This situation is reminiscent of the
Altarelli-Feruglio model \cite{Altarelli:2005yp,Altarelli:2005yx}
where one $\mathbb Z_2$ is contained in $A_4$ while the other $\mathbb
Z_2$, corresponding to a $\mu$-$\tau$ symmetry, arises
accidentally. In the case of $T_7$, none of the two $\mathbb Z_2$
factors of the tribimaximal Klein symmetry is a part of the
underlying family symmetry. As a consequence, one expects no
protection of the tribimaximal structure when NLO terms are taken into account.

\subsection{Next-to-leading order effects}

The next-to-leading order superpotential for the charged and neutral leptons contains all terms up to mass dimension six and seven, respectively, that are invariant under the gauge and family symmetries. Here, we only indicate the terms that are not already part of the leading-order superpotential:
\begin{align}
\Delta W_{\ell} =\, \frac{1}{\Lambda^2} \bigg( &\Cten{} \, L \, \ec{} \, \hd{} \, \varphi \, \varphi \, + \, \Celeven{} \, L \, \ec{} \, \hd{} \, \varphi \, \phit{} \, + \, \Ctwelve{} \, L \, \ec{} \, \hd{} \, \phit{} \, \phit{} \, + \nonumber\\
& \Cfifteen{} \, L \, \mu^c \, \hd{} \, \varphi \, \varphi \, + \, \Csixteen{} \, L \, \mu^c \, \hd{} \, \varphi \, \phit{} \, + \, \Cseventeen{} \, L \, \mu^c \, \hd{} \, \phit{} \, \phit{} \, + \label{eq:NLO-charged-leptons}\\
& \Ctwenty{} \, L \, \tau^c \, \hd{} \, \varphi \, \varphi \, + \, \Ctwentyone{} \, L \, \tau^c \, \hd{} \, \varphi \, \phit{} \, + \, \Ctwentytwo{} \, L \, \tau^c \, \hd{} \, \phit{} \, \phit{} \bigg),\nonumber\\
\nonumber
\Delta W_{\nu} =\, \frac{1}{\Lambda^3} \bigg( &\Cthree{} \, (L \, L )_{\bs{3}}\, \hu{} \, \hu{} \, \varphi \, \varphi \, + \, \Cfour{} \, (L \, L)_{\bs{\overline{3}}} \, \hu{} \, \hu{} \, \varphi \, \varphi \, \, + \Cseven{} \, (L \, L)_{\bs{3}} \, \hu{} \, \hu{} \, \phit{} \, \phit{} \, + \nonumber\\
&\Ceight{} \, (L \, L)_{\bs{\overline{3}}} \, \hu{} \, \hu{} \, \phit{} \, \phit{} \, \, + \Cfive{} \, (L \, L)_{\bs{3}} \, \hu{} \, \hu{} \, \varphi \, \phit{} \, + \, \Csix{} \, (L \, L)_{\bs{\overline{3}}} \, \hu{} \, \hu{} \, \varphi \, \phit{} \bigg). \label{eq:NLO-neutral-leptons} 
\end{align}

The subscripts in \ref{eq:NLO-neutral-leptons} indicate that we have contracted the family indices of the corresponding product to obtain a $\bs{3}$ and $\bs{\overline{3}}$, respectively. As before, contracting the family and gauge indices, and substituting the \vevs{} for the Higgs and flavon fields gives:
\begin{align}
\Delta M_{\ell} &=  \frac{v_{d}}{\Lambda^2} \, \frac{1}{3} 
\bigg[ \Cten{}\left(\begin{smallmatrix}0 & 0 & 0\\0 & 0 & 0\\0 & 0 & 0\end{smallmatrix}\right) {v_{\varphi}^2} +  \Celeven{}  \left(\begin{smallmatrix}1 & 0 & 0\\1 & 0 & 0\\1 & 0 & 0\end{smallmatrix}\right) \frac{v_{\varphi}}{\sqrt{3}}v_{\wt \varphi} +  \Ctwelve{}  \left(\begin{smallmatrix}1 & 0 & 0\\0 & 0 & 0\\0 & 0 & 0\end{smallmatrix}\right) v_{\wt \varphi}^2 \notag\\
&+  \Cfifteen{}\left(\begin{smallmatrix}0 & 0 & 0\\0 & 0 & 0\\0 & 0 & 0\end{smallmatrix}\right) {v_{\varphi}^2} +  \Csixteen{}  \left(\begin{smallmatrix}0 & 1 & 0\\0 & 1 & 0\\0 & 1 & 0\end{smallmatrix}\right) \frac{v_{\varphi}}{\sqrt{3}}v_{\wt \varphi} +  \Cseventeen{}  \left(\begin{smallmatrix}0 & 0 & 0\\0 & 1 & 0\\0 & 0 & 0\end{smallmatrix}\right) v_{\wt \varphi}^2 \label{eq:delta_M_ell}\\
&+  \Ctwenty{}\left(\begin{smallmatrix}0 & 0 & 0\\0 & 0 & 0\\0 & 0 & 0\end{smallmatrix}\right) {v_{\varphi}^2} +  \Ctwentyone{} \left(\begin{smallmatrix}0 & 0 & 1\\0 & 0 & 1\\0 & 0 & 1\end{smallmatrix}\right) \frac{v_{\varphi}}{\sqrt{3}}v_{\wt \varphi} +  \Ctwentytwo{} \left(\begin{smallmatrix}0 & 0 & 0\\0 & 0 & 0\\0 & 0 & 1\end{smallmatrix}\right) v_{\wt \varphi}^2 \bigg],\notag\\[2ex]
\Delta M_{\nu} &= \frac{v_{u}^{2}}{\Lambda^3} \frac{1}{9\sqrt{2}} \bigg[ \Cthree{}\left(\begin{smallmatrix}0 & 0 & 0\\0 & 0 & 0\\0 & 0 & 0\end{smallmatrix}\right) v_{\varphi}^2 +  \sqrt{3} \Cfour{}  \left(\begin{smallmatrix}2 & -\omega^2 & -\omega\\-\omega^2 & 2\omega & -1\\-\omega & -1 & 2\omega^2\end{smallmatrix}\right) v_{\varphi}^2 +  \sqrt{6} \Cseven{} \left(\begin{smallmatrix}1 & 0 & 0\\0 & 0 & 1\\0 & 1 & 0\end{smallmatrix}\right) v_{\wt \varphi}^2 \notag\\
&+  \sqrt{6} \Ceight{} \left(\begin{smallmatrix}2 & 0 & 0\\0 & 0 & - 1\\0 & - 1 & 0\end{smallmatrix}\right) v_{\wt \varphi}^2 +  \sqrt{2} \Cfive{} \left(\begin{smallmatrix}1 & \omega & \omega^2\\\omega & \omega^2 & 1\\\omega^2 & 1 & \omega\end{smallmatrix}\right) v_{\varphi} v_{\wt \varphi} + \Csix{} \left(\begin{smallmatrix}2 & -\omega & -\omega^2\\-\omega & 2\omega^2 & -1\\-\omega^2 & -1 & 2\omega\end{smallmatrix}\right) v_{\varphi} v_{\wt \varphi} \bigg]. \label{eq:delta_M_nu}
\end{align}

A quick inspection of \ref{eq:delta_M_ell} shows that the contributions proportional to $\Cten{}$, $\Cfifteen{}$, $\Ctwenty{}$ are identically zero and have no effect whatsoever on the mass matrices. The contributions proportional to $\Ctwelve{}$, $\Cseventeen{}$, $\Ctwentytwo{}$ do not disturb the diagonal structure of $M_{\ell}$ and hence do not affect the bi-unitary transformations that diagonalize it and that enter in the definition of \UPMNS{}. In contrast, $\Celeven{}$, $\Csixteen{}$, $\Ctwentyone{}$ will lead to a departure from tribimaximal mixing, where it has to be noted that the magnitudes of the coefficients have to be weighted by the corresponding Yukawa couplings, and hence the effect of $\Celeven{}$ will be negligible.

\medskip

From \ref{eq:delta_M_nu} we see that the term proportional to $\Cthree{}$ is identically zero and will not have an impact on the mixing matrix. The term corresponding to $\Cseven{}$ is non-vanishing, but commutes with the two generators of the tribimaximal Klein symmetry (see e.g.~\cite{Altarelli:2010gt} for the explicit form of the generators), and as a consequence does not disturb the tribimaximal form of the mixing matrix.

\medskip

The NLO terms that do cause a departure from tribimaximal mixing are those
corresponding to $\Cfour{}$, $\Ceight{}$, $\Cfive{}$, $\Csix{}$. The
contribution proportional to  $\Ceight{}$ is invariant under a $\mu-\tau$
symmetry (which corresponds to one of the generators of the tribimaximal Klein
symmetry) and thus leaves $\theta_{23}$ and $\theta_{13}$ unchanged. Of
particular interest is the term proportional to $\Cfive{}$, since its effect
is such that $\theta_{12}$ is almost unchanged, $\theta_{23}$ stays within its
$3\sigma$ interval, and $\theta_{13}$ receives a sizable
contribution. Therefore it seems to be a good strategy to try and suppress all
NLO contributions to the superpotential except the one proportional to
$\Cfive{}$. Thus, we will keep the deviation of $\theta_{12}$ and
$\theta_{23}$ from the tribimaximal case small, and at the same time achieve a
non-vanishing $\theta_{13}$ in agreement with experiment. In \ref{sec:uv-mo}
we present a renormalizable, ultraviolet completion of our model that meets
exactly these criteria outlined above. Before that, however, we will explore
in \ref{sec:numreslt} the phenomenology of our model at the effective level. We first consider the
model in its most general form without any specific assumptions about the
coefficients. At the end of~\ref{sec:numreslt} we also discuss the case where
all coefficients which break the tribimaximal structure, except for
$\Cfive{}$, are set to zero.

\subsection{Phenomenology}
\label{sec:numreslt}

We now compare our results with experiment \cite{An:2012eh,Ahn:2012nd,Fogli:2012ua,Tortola:2012te,GonzalezGarcia:2012sz}. To that end, we first have to specify the coefficients in the leading and next-to-leading order superpotential (cf.~\ref{eq:leading-order-superpotential-ell}, \ref{eq:leading-order-superpotential-nu}, \ref{eq:NLO-charged-leptons}, \ref{eq:NLO-neutral-leptons}). The mixing angles, CP phases and masses will then be uniquely determined.

\medskip

\begin{figure}[h!]
\centering
\includegraphics[width=0.65\textwidth]{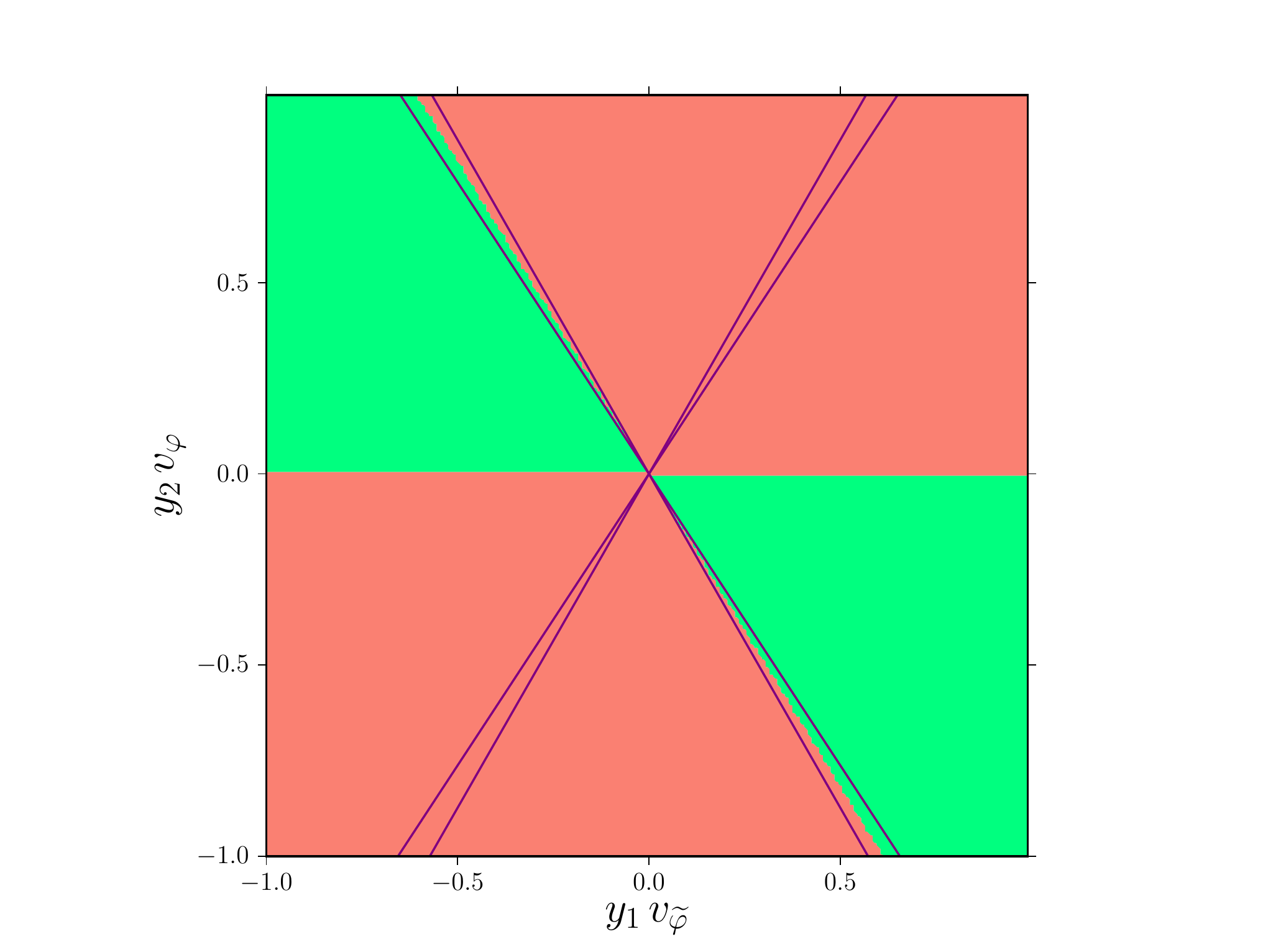}
\setcapindent{0em}
\caption{Contour lines of $\Delta m_{31}^2/\Delta m_{21}^2 \equiv 30$ as a
  function of $y_1 v_{\phit}$ and $y_2 v_{\varphi}$ for normal neutrino mass ordering. The green region corresponds to values of the parameters where the mixing angles are tribimaximal.
}
\label{fig:LO-mass-ratios}
\end{figure}
 
The leading-order superpotential gives tribimaximal mixing which is form-diagonalizable \cite{Low:2003dz}. As such, the mixing angles are in principle independent of the coefficients \Cnine{}, \Cfourteen{}, \Cnineteen{}, \Ctwo{}, \Cone{}. However, changing the coefficients affects the neutrino masses and can lead to a reordering of the mass eigenstates that in turn has an impact on the mixing angles. Since we work in a basis where the charged lepton mass matrix is diagonal, we can identify \Cnine{}, \Cfourteen{}, \Cnineteen{} with the corresponding Yukawa couplings for which we substitute their experimentally determined values \cite{Beringer:2012}. For \Ctwo{} and \Cone{}, we choose values such that we can fit the mass ratio $\Delta m_{31}^2/\Delta m_{21}^2$. 

\medskip

In \ref{fig:LO-mass-ratios} we present the contour lines of $\Delta
m_{31}^2/\Delta m_{21}^2\equiv 30$ as a function of $y_1 v_{\phit}$
and $y_2 v_{\varphi}$ (cf.~\vref{eq:LOneutrinomassmatrix}) for normal
neutrino mass orderings. Here we do not consider the case of inverted
neutrino mass orderings as it would require $y_1 v_{\phit} \ll y_2
v_{\varphi}$.  The green region in \ref{fig:LO-mass-ratios}
corresponds to values of the parameters for which we obtain
tribimaximal mixing, whereas in the red region, the mixing matrix
corresponds to a permutation of the columns of \UPMNS{},
i.e.~tribimaximal mixing and the desired mass hierarchy cannot be
simultaneously satisfied. Note that we have chosen $y_1$ and $y_2$ to
be real for presentational purposes only. In the general case, $y_1$
and $y_2$ are complex, and to fit the mass ratio is equally easy.

\medskip

In the next-to-leading order superpotential, we assume that the fundamental scale of the theory is about one order of magnitude larger than the flavon \vevs{} and fix the ratio at $v_{\varphi}/\Lambda = v_{\phit}/\Lambda = 0.10$ (see \ref{eq:delta_M_ell} and \ref{eq:delta_M_nu}). For the coefficients $\Cthree{},\ldots,\Ctwentytwo{}$, we choose 1000 sets of complex numbers  whose real and imaginary parts are uniformly distributed between -2.5 and +2.5. For each such set, we calculate the corrections to the charged lepton and neutrino mass matrices and recalculate the mixing angles and masses. Note that now, $M_{\ell}$ is not necessarily diagonal, and we have to resort to the general definition of \UPMNS{} in terms of the bi-unitary transformations that diagonalize the charged lepton and neutrino mass matrices. 

\medskip

\begin{figure}[h!]
\centering
\subfigure[\footnotesize $v_{\varphi}/\Lambda = v_{\phit}/\Lambda = 0.10$ and $-2.5\leq \mathfrak{Re\,}C^\alpha_i,\, \mathfrak{Im\,}C^\alpha_i \leq2.5$.]{
\includegraphics[width=1\textwidth]{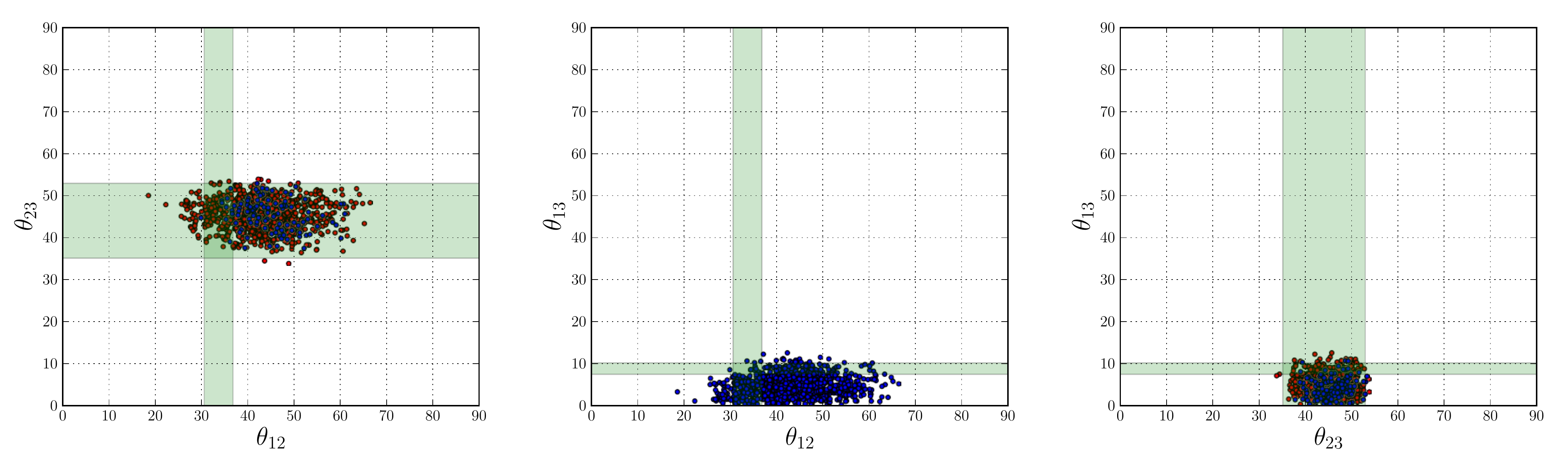}
\label{fig:scatter1-a}
}
\newline
\subfigure[\footnotesize $v_{\varphi}/\Lambda = 0.25, \enspace v_{\phit}/\Lambda = 0.05, \enspace -2\leq \mathfrak{Re\,}\Cfive{},\, \mathfrak{Im\,}\Cfive{} \leq 2$,\, all other coefficients set to zero.]{
\includegraphics[width=1\textwidth]{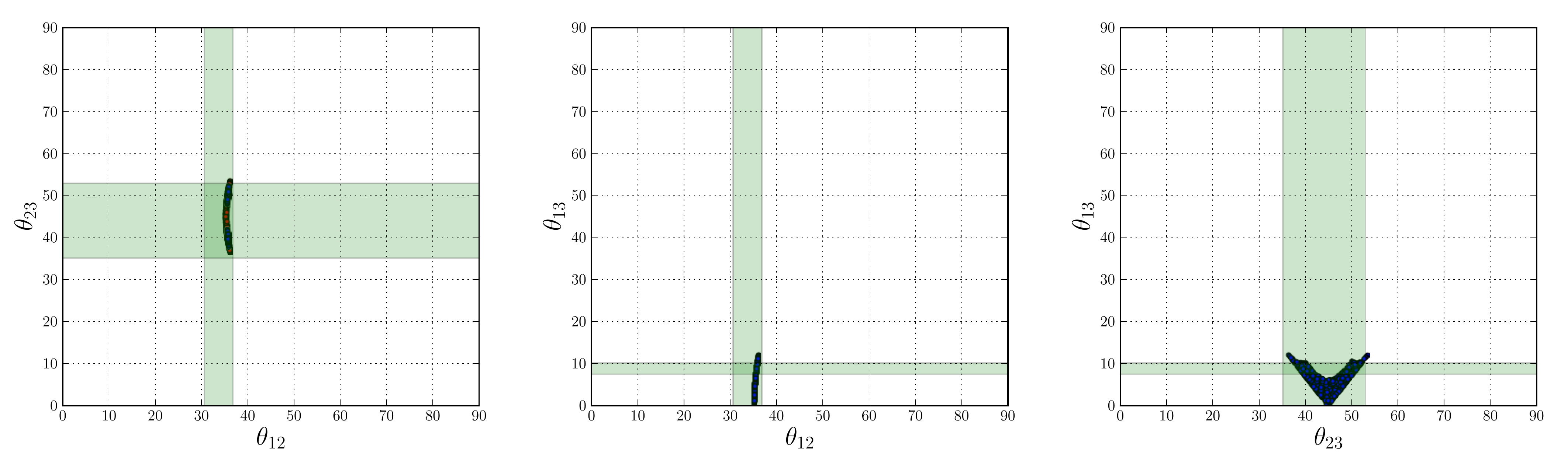}
\label{fig:scatter1-b}
}
\setcapindent{0em}
\caption{The mixing angles for a given set of leading order coefficients $\Ctwo{}$, $\Cone{}$, $\Cnine{}$, $\Cfourteen{}$, $\Cnineteen{}$ and 1000 sets of random values for the next-to-leading order coefficients $\Cthree{},\ldots,\Ctwentytwo{}$. Our choice for the coefficients $y_1=0.50$, $y_2= -1.52\, y_1\, v_{\tilde\varphi}/v_{\varphi}$ correspond to $\Delta m_{31}^2/\Delta m_{21}^2\sim30$ in the LO approximation. The green bands correspond to the $3\sigma$ intervals \cite{Fogli:2012ua} for the mixing angles (normal hierarchy in the case of $\theta_{23}$ and $\theta_{13}$). In each diagram, the color of the markers indicates whether the data point lies in the $3\sigma$ interval of the respective angle that is not plotted (blue) or not (red).
}
\label{fig:scatter1}
\end{figure}

We present the results in \vref{fig:scatter1-a}. The LO order prediction has given way to a ``blob'' centered around the tribimaximal values for the mixing angles. Remarkably, 99\% of the points lie in the $3\sigma$ interval \cite{Fogli:2012ua} of $\theta_{23}$, whereas 13\% also satisfy the $3\sigma$ bounds for $\theta_{12}$ (see first panel of \ref{fig:scatter1-a}). As was to be expected, the strongest constraint comes from $\theta_{13}$. From the third panel of \ref{fig:scatter1-a} we see that low values for $\theta_{13}$ are preferred, but 10\% of the points can accommodate a large reactor angle in agreement with experiment. This number drops to 1\% when we take the constraints from the other two angles into account.

\medskip

In \vref{fig:scatter1-b} we present the results for a model where all terms except \Cfive{} that lead to departures from tribimaximal mixing are suppressed. This will correspond to the case of the renormalizable model that we will discuss in \ref{sec:uv-mo}. As we can see from the first panel of \ref{fig:scatter1-b}, $\theta_{12}$ is practically unchanged from its tribimaximal value, whereas $\theta_{23}$ varies very little, and both angles are well within their $3\sigma$ error bands. At the same time, we can easily obtain a large $\theta_{13}$ in such a way that the experimental constraints on the other two angles are always satisfied (blue points in the third panel of \ref{fig:scatter1-b}). At the same time we can observe an interesting correlation between $\theta_{23}$ and $\theta_{13}$, namely the larger $\theta_{13}$ is, the more $\theta_{23}$ deviates from maximal mixing.\footnote{This numerical observation is a reflection of a well-known mixing sum rule relating the deviations of the atmospheric and reactor angles from their tribimaximal values via the CP phase $\delta$~\cite{King:2009qt,King:2011zj}.} Unfortunately, this effect is symmetric around $\theta_{23}=45^\circ$ and hence does not give a hint whether this deviation is positive or negative.

\medskip

\section{An ultraviolet completion of the model}
\label{sec:uv-mo}

We have shown in \ref{sec:TBMLO} how the minimal $T_7$ model, which predicts exact
tribimaximal lepton mixing at leading order, receives in general significant corrections at
next-to-leading order. Such deviations from the tribimaximal pattern are
welcome in order to accommodate the observed large value of the reactor angle
$\theta_{13}$.  On the other hand, the same set of corrections can potentially modify
the other two mixing angles to values disfavored by current global fits. In
particular, the solar angle should obtain only small corrections, as its
tribimaximal value of $\theta_{12}\approx 35.3^\circ$ fits the experimental
result already remarkably well. In order to obtain a deviation from
tribimaximal mixing which stabilizes the solar angle in first approximation, one can
make use of corrections of the so-called \textit{trimaximal}
type~\cite{Haba:2006dz,He:2006qd,Lam:2006wm,Grimus:2008tt,Albright:2008rp,Albright:2010ap,Ishimori:2010fs,Shimizu:2011xg,He:2011gb}. These are typically defined by corrections to the tribimaximal
neutrino mass matrix whose eigenvectors are proportional to either $(1,1,1)^T$~\cite{King:2011zj} or
$(2,-1,-1)^T$~\cite{Antusch:2011ic}. Comparison with \ref{eq:delta_M_nu} reveals
that there exists exactly one NLO correction which respects the trimaximal
structure but breaks the tribimaximal one. This subleading contribution to the
neutrino mass matrix is proportional to the coefficient $\Cfive$ and can be
traced back to an NLO operator in \ref{eq:NLO-neutral-leptons} in which $LL$
is contracted to a ${\bf 3}$ of $T_7$. Notice that the other two terms with
$LL$ contracted to a ${\bf 3}$, i.e. those proportional to $\Cthree$ and
$\Cseven$, do not lead to deviations from the tribimaximal structure.

This begs the question of how to remove or suppress the unwanted NLO
operators while keeping the one proportional to $\Cfive$. 
To this end, it is useful to recall that we have so far only discussed the
model at the effective level, i.e.~writing down all (non-renormalizable)
terms which are allowed by the imposed symmetries. However, any particular ultraviolet completion of such an
effective model does not give rise to all effective NLO terms. The underlying
renormalizable model will involve messenger fields which have to be introduced
to mediate the effective leading order operators. These necessary messengers
typically do not generate all NLO terms \cite{Varzielas:2010mp}. If one particular NLO term is
desired, as is the case in our $T_7$ model, an extra messenger has to be
introduced.  In the following we adopt the strategy of considering an
ultraviolet completion of the effective $T_7$ model such that the operator
proportional to $\Cfive$ is allowed, while the bothersome NLO terms are forbidden or
sufficiently suppressed. We  begin the discussion of the ultraviolet
completion of the model with the charged lepton sector, before turning to the
more interesting neutrino sector which will be responsible for the breaking of
the tribimaximal to the trimaximal pattern.

\subsection{\label{sec:UV-ell}The charged lepton sector}

The leading order operators in \ref{eq:leading-order-superpotential-ell} can
be obtained by imposing a $T_7$ triplet messenger pair $\Theta$, $\Theta^c$
with the transformation properties as given in \ref{tab:uv-model}.
\begin{table}[t]
\begin{center}
\begin{tabular}{|c||c||c|c||c|c|c|c|}\hline
$\phantom{\Big|}$Field$\phantom{\Big|}$ &
$\Delta$ 
& $\Theta$ & $\Theta^c$
& $\Sigma$ &$ \Sigma^c$& $\Omega$ & $\Omega^c$ 
  \\\hline
$\phantom{\Big|}$$T_7$$\phantom{\Big|}$ &  ${\bf  1}$
& ${\bf 3}$ & ${\bf \ol 3}$
& ${\bf 3}$ & ${\bf \ol 3}$ & ${\bf 3}$ & ${\bf \ol 3}$ \\\hline
$\phantom{\Big|}$$\U{1}_Y$ $\phantom{\Big|}$&   $2$ 
& $-2$ & $2$
 & $-2$ & $2$& $1$ & $-1$
 \\\hline
$\phantom{\Big|}$$\U{1}_R$$\phantom{\Big|}$ & $0$ 
& $1$ & $1$
 & $2$ & $0$& $1$ & $1$
 \\\hline
\end{tabular}
\end{center}
\setcapindent{0em}
\caption{\label{tab:uv-model}The charge assignments of the (effective) Higgs
  $\SU{2}_L$ triplet $\Delta$ as well as of the messenger fields required for the
  ultraviolet completion of the minimal $T_7$ model.}
\end{table}
The resulting renormalizable charged lepton superpotential takes the form
\be
W_\ell^{\mathrm{ren}} ~\sim~ 
L\hd{}\Theta^c + \Theta \wt\varphi \ec + \Theta \wt\varphi \muc + \Theta
\wt\varphi \tauc
+\Theta \Theta^c (M_{\Theta}+ \varphi + \wt\varphi)  \ ,
\ee
where we have suppressed all dimensionless coupling coefficients. Note
that $\Theta$ and $\Theta^c$ allow for a bilinear mass term. However,
due to the absence of any shaping symmetry, we can also couple a
flavon triplet to this product, thus leading to a trilinear term
which, after family symmetry breaking, gives a correction to the
messenger mass. This is a general feature for heavy messenger fields
which transform as triplets under $T_7$. Ignoring the correction to
the messenger mass, we can integrate out the pair $\Theta$, $\Theta^c$
and obtain the effective superpotential of
\ref{eq:leading-order-superpotential-ell} without any higher order
corrections. Hence, in the present ultraviolet completion, the only
source of deviations from a diagonal charged lepton mass matrix
originates from the aforementioned trilinear terms correcting the
messenger mass. Such a correction can be rendered sufficiently small
by assuming $\vevof{\wt\varphi},\vevof{\varphi} \ll M_\Theta$. In the
following, we will therefore ignore any NLO terms in the charged
lepton sector.

\subsection{\label{sec:UV-nu}The neutrino sector}

We now turn to the neutrino sector where we want to formulate a
renormalizable theory which, at leading order, gives rise to the
effective superpotential in
\ref{eq:leading-order-superpotential-nu}. The most popular possibility
to derive the Weinberg operator from renormalizable terms is provided
by the famous type~I seesaw
mechanism~\cite{Minkowski:1977sc,Gell-Mann,Yanagida,Mohapatra:1979ia}. However,
applied to the minimal $T_7$ model without any shaping symmetry, the
type~I seesaw mechanism does not yield the tribimaximal structure at
leading order. This can be traced back to the fact that the triplet
representations of $T_7$ are complex. Putting the $L$ into the ${\bf
3}$ would entail to introduce right-handed neutrinos $N^c$ in the
${\bf \ol 3}$ so that the Dirac neutrino Yukawa term $L\hu{}N^c$ leads
to a Dirac mass matrix that is proportional to the identity matrix. Then the
tribimaximal structure of the effective light neutrino mass matrix
would be inherited directly from a tribimaximal right-handed neutrino
mass matrix $M_R$. With $N^c$ being a ${\bf \ol 3}$ rather than a
${\bf 3}$, the alignments of the ${\bf \ol 3}$ flavon $\wt\varphi$ and
the ${\bf 3}$ flavon $\varphi$ would have to be exchanged to get $M_R$
of tribimaximal form. However, changing the alignment of the flavon
$\wt\varphi$ coupling to the charged leptons to $(1,1,1)^T$ destroys
the diagonal charged lepton mass matrix in~\ref{eq:mellLO}. As a
result, under the fairly general set of assumptions that we made
above, the total leading order lepton mixing would no longer feature
the tribimaximal pattern.

\medskip

As an alternative to the type~I seesaw mechanism, we can adopt the type~II
seesaw~\cite{Magg:1980ut,Schechter:1980gr,Wetterich:1981bx,Lazarides:1980nt,Mohapatra:1980yp}. 
For the sake of clarity, we assume a Higgs $\SU{2}_L$ triplet
$\Delta$ in the remainder of this section. It is straightforward to replace
such a Higgs field by the product of two $\hu{}$ doublets which couple to the square of
the lepton doublet $L$ via an $\SU{2}_L$ triplet messenger field. The type~II seesaw
mechanism seems particularly suited for our purposes as we wish to obtain the
NLO correction to the neutrino mass matrix proportional to the coupling constant
$\Cfive$, which arises from contracting $LL$ to a ${\bf 3}$ of $T_7$. This can
be naturally achieved by invoking a messenger field $\Sigma^c$ in the ${\bf
  \ol 3}$ representation which leads to the renormalizable term
$LL\Sigma^c$. The $\U{1}_Y$ and $\U{1}_R$ charges required for the existence of
this term are given in~\ref{tab:uv-model}. Demanding a bilinear mass term
fixes the charge assignments of $\Sigma$ which, in turn, can couple to the 
Higgs field $\Delta$ and the flavon~$\wt\varphi$. A similar coupling to the other
flavon $\varphi$ is forbidden by the $T_7$  family symmetry. We are therefore
forced to introduce another pair of messengers $\Omega$, $\Omega^c$, which
allows to generate the second leading order contribution to $M_\nu$
proportional to~$\Cone$, cf.~\ref{eq:mnuLO}. Their charge assignments are again
given in \ref{tab:uv-model}.

\medskip

\begin{figure}[t]
\begin{center}
\includegraphics[height=3cm]{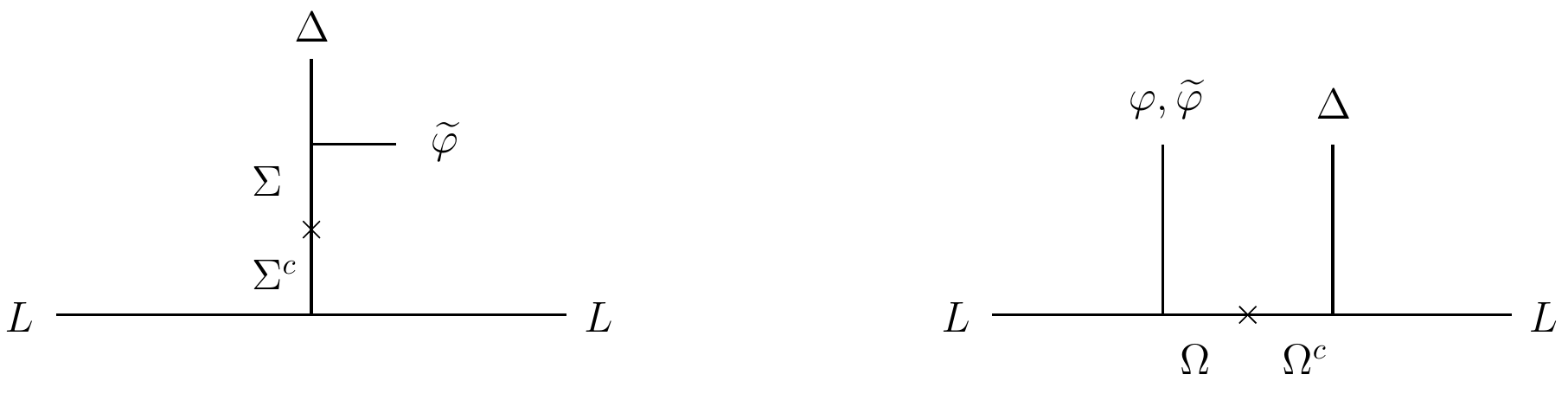}
\end{center}
\setcapindent{0em}
\caption{\label{fig:LO-dia}The diagrams contributing to the leading order
  neutrino superpotential of~\ref{eq:leading-order-superpotential-nu}.}
\end{figure}

With this particle content, the renormalizable neutrino superpotential reads
\bea
W^{\mathrm{ren}}_\nu &\sim& 
 LL\Sigma^c + \Sigma \wt \varphi \Delta 
+ \Sigma \Sigma^c (M_\Sigma + \varphi +\wt \varphi) \notag\\
&&+\, L\varphi \Omega + L \wt \varphi \Omega + \Omega^c \Delta L 
+ \Omega \Omega^c (M_\Omega + \varphi +\wt \varphi) \label{eq:ren-NU} \\
&& +\, L \Sigma^c \Omega^c + \Sigma^c\Omega^c\Omega^c \ , \notag
\eea
where we have suppressed all dimensionless coupling constants.
As already discussed for the charged lepton sector, the masses of the $T_7$ triplet messengers
receive corrections from the flavon \vevs{}. These will break the tribimaximal
structure at higher order. Ignoring the corrections to the messenger masses as
well as the two operators in the third line of~\ref{eq:ren-NU} results in the
effective superpotential
of~\ref{eq:leading-order-superpotential-nu}. The corresponding diagrams are
sketched in~\ref{fig:LO-dia}, where the right diagram contributes to
the term proportional to $\Cone$, while the term proportional to $\Ctwo$
arises from both diagrams. 

\medskip

\begin{figure}[t]
\begin{center}
\includegraphics[height=3cm]{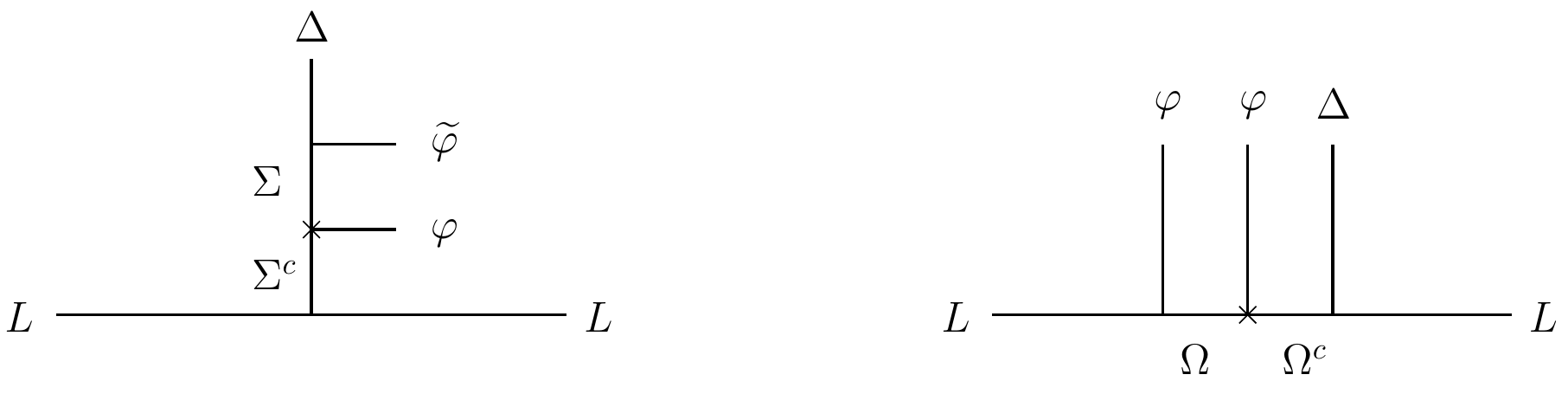}
\end{center}
\setcapindent{0em}
\caption{\label{fig:NLO-dia}Two diagrams contributing to the NLO 
  neutrino superpotential. The diagram on the left gives the desired
  trimaximal correction, while the diagram on the right should be suppressed
  in order to avoid large deviations of the solar angle from its
  tribimaximal value of $\theta_{12}\approx 35.3^\circ$.}
\end{figure}

We point out that the left diagram in \ref{fig:LO-dia} is needed only for generating the desired trimaximal NLO correction while suppressing
other unwanted NLO terms. In other words, if we wanted to end up with
tribimaximal neutrino mixing, the diagram with the $\Omega$ messenger would
have been perfectly sufficient. In that case we would have to impose the condition
$\vevof{\wt\varphi},\vevof{\varphi} \ll M_\Omega$, similar to the situation in
the charged lepton sector. However, since accurate tribimaximal mixing is
ruled out by the Daya Bay and Reno measurements, we are forced to consider
{\it sizable} NLO corrections. Due to the structure of the diagram with the $\Omega$
messenger, the NLO terms obtained from attaching a flavon to the cross
representing the mass term $M_\Omega \Omega \Omega^c$ would give rise to
corrections that would shift the solar angle away from its experimentally
allowed region. Hence, the trimaximal NLO correction must originate from
another diagram, namely the one with the $\Sigma$ messenger, more precisely,
the left diagram of~\ref{fig:LO-dia} with the flavon $\varphi$ attached to the
cross. The resulting diagram is sketched in~\ref{fig:NLO-dia}, where one NLO
diagram involving the $\Omega$ messenger is also shown. 
The NLO diagram on the right can be suppressed compared to the NLO diagram on
the left by assuming a hierarchy in the messenger masses. This can be
parameterized by
\be
M_\Sigma \sim \epsilon^k M_\Omega \ ,
\ee
where $k$ is a positive integer and $\epsilon$ an expansion parameter around
$0.2$. With this hierarchy of messenger masses imposed we also have to assume
a hierarchy in the two flavons \vevs{},
\be
\frac{v_{\wt\varphi}}{v_{\varphi}} \sim \frac{M_\Sigma}{M_\Omega} \sim 
\epsilon^k \ ,
\ee
so that the leading order mass contributions derived from the two diagrams
in~\ref{fig:LO-dia} are of similar size. Note that the diagram on the right
involving the flavon $\wt\varphi$ gives a subdominant contribution which is of
the same tribimaximal structure as the one obtained form the left diagram.
Finally, we need to guarantee that the trimaximal NLO correction of the left
diagram in~\ref{fig:NLO-dia} is sizable enough to account for the observed
value of reactor angle $\theta_{13}$. This requires the correction to the
mass of the $\Sigma$ messenger originating from the flavon $\varphi$ to be of
order $\epsilon$, i.e.
\be
\frac{v_\varphi}{M_\Sigma} \sim \epsilon \ .
\ee
Defining $M \sim M_\Omega$ we can summarize the requirements on the
hierarchies as
\be
v_{\wt\varphi} \sim \epsilon^{2k+1} M \ , ~\quad
v_{\varphi} \sim \epsilon^{k+1} M \ , ~\quad
M_\Sigma \sim \epsilon^k M \ ,~\quad
M_\Omega \sim M \ ,
\ee
yielding neutrino mass contributions proportional to $\frac{v_{\wt\varphi}}{M_\Sigma} \sim
\frac{v_\varphi}{M_\Omega} \sim \epsilon^{k+1}$ at leading order, with a
trimaximal NLO correction proportional to $\epsilon^{k+2}$. In comparison, the right
diagram in~\ref{fig:NLO-dia} gives a contribution proportional to $\epsilon^{2k+2}$
which is suppressed by a factor of
$\epsilon^k$ relative to the desired NLO correction. 

\medskip

With the above  assumptions on the hierarchies it is possible to suppress all
unwanted contractions in~\ref{eq:NLO-neutral-leptons} while keeping the
desired one proportional to $\Cfive$. This holds true even if we take into
consideration possible higher order diagrams obtained from the two terms of
the third line of~\ref{eq:ren-NU}. These terms can give rise to diagrams which
involve one $\Sigma$ messenger as well as one or two $\Omega$
messengers. Multiplying the appropriate flavon fields, one can easily find
that the maximal contribution derived from the renormalizable term
$L\Sigma^c\Omega^c$ is of order $\epsilon^{2k+2}$, while the one derived from
$\Sigma^c\Omega^c\Omega^c$ is of order $\epsilon^{3k+3}$. This shows that these
higher order corrections are also suppressed with respect to the trimaximal NLO correction by at
least a factor of $\epsilon^k$. 

\medskip

We conclude this section by mentioning that we have also checked these results
numerically for the case of $k=2$. Integrating out the messengers we have
first determined the effective neutrino mass matrix, from which we have
calculated the mixing angles using the Mixing Parameter Tools provided with
the REAP package\cite{Antusch:2005gp}. This numerical check confirms the 
results of this section, showing how an ultraviolet completion of the minimal
$T_7$ model can lead to deviations from tribimaximal mixing which are
compatible with a sizable reactor angle as well as a solar angle that is close
to its tribimaximal value.

\section{Vacuum alignment}
\label{sec:vev-4}

\subsection{General discussion}
\label{sec:vevgd}

Any model of flavor that tries to explain a particular mixing pattern by means
of a family symmetry which gets broken  by flavon \vevs{} has to justify the
imposed alignments. In this section we study the possibilities of achieving
the vacuum configuration required by the minimal $T_7$ model.
Following the idea of \cite{Altarelli:2005yx}, we demand that supersymmetry
be unbroken at the scale of family symmetry breaking. This so-called $F$-term
alignment mechanism introduces driving fields whose $F$-terms are set to zero,
thus entailing the important $F$-term conditions. Just like the regular
flavons, driving fields are neutral under the $\SU{2}_L \times \U{1}_Y$ gauge symmetry.
With the five irreducible representations of $T_7$ and no $\mathbb Z_N$ shaping
symmetry, one can introduce driving fields in at most five different representations. 
Hence, we obtain five distinct $F$-term conditions which we discuss in the
following. Throughout this section we will drop all dimensionless coupling
as well as overall CG coefficients.

\medskip

A driving field $D_{\bf 1}$ transforming in the ${\bf 1}$ of $T_7$ allows for two
invariant terms in the renormalizable flavon superpotential
\begin{equation}
W_{\mathrm{flav}} ~ \supset ~D_{\bf 1} \left( M^2 +  \wt \varphi \varphi \right) \ ,
\end{equation}
which lead to the $F$-term condition
\be
\frac{\partial W_{\mathrm{flav}}} {  \partial D_{\bf 1}}=M^2+\langle \wt \varphi_1 \rangle\langle  \varphi_1 \rangle+\langle \wt \varphi_2 \rangle\langle  \varphi_2 \rangle+\langle \wt \varphi_3 \rangle\langle  \varphi_3 \rangle ~=~0\ .
\ee
Similarly, the driving fields $D_{\bf{r}}$ in the representations ${\bf{r}} =
{\bf 1'}, {\bf 1''}, {\bf 3}, {\bf \ol 3}$ of $T_7$ lead to the conditions:
{\small
\begin{eqnarray}
\frac{\partial W_{\mathrm{flav}}} {  \partial D_{\bf 1'}}&\!=\!&\langle \wt \varphi_1 \rangle\langle \varphi_2 \rangle+\langle \wt \varphi_2 \rangle\langle \varphi_3 \rangle+\langle \wt \varphi_3 \rangle\langle \varphi_1 \rangle ~=~0
\ , \label{eq:F1}\\ 
\frac{\partial W_{\mathrm{flav}}} {  \partial D_{\bf 1''}}&\!=\!&\langle \wt \varphi_1 \rangle \langle \varphi_3 \rangle+\langle \wt \varphi_2 \rangle \langle \varphi_1 \rangle+\langle \wt \varphi_3 \rangle\langle \varphi_2 \rangle ~=~0
\ ,\label{eq:F2} \\
\nonumber\\
\frac{\partial W_{\mathrm{flav}}} {  \partial D_{\bf 3}}&\!=\!&M_1\begin{pmatrix} \langle \wt \varphi_1 \rangle\\\langle \wt \varphi_2 \rangle\\\langle \wt \varphi_3 \rangle\end{pmatrix}+
\begin{pmatrix} 
\langle \wt \varphi_1 \rangle^2+2\langle \wt \varphi_2 \rangle\langle \wt \varphi_3 \rangle\\\omega^2(\langle \wt \varphi_3 \rangle^2+2\langle \wt \varphi_1 \rangle\langle \wt \varphi_2 \rangle)\\\omega(\langle \wt \varphi_2 \rangle^2+2\langle \wt \varphi_3 \rangle\langle \wt \varphi_1 \rangle)
\end{pmatrix}\nonumber\\
&&+\begin{pmatrix} 
\langle \varphi_1 \rangle^2-\langle \varphi_2 \rangle\langle \varphi_3 \rangle\\\langle \varphi_2 \rangle^2-\langle \varphi_3 \rangle\langle \varphi_1 \rangle\\\langle \varphi_3 \rangle^2-\langle \varphi_1 \rangle\langle \varphi_2 \rangle
\end{pmatrix}+
\begin{pmatrix} 
\langle \wt \varphi_1 \rangle\langle \varphi_1 \rangle+\omega \langle \wt \varphi_2 \rangle\langle \varphi_2 \rangle+\omega^2 \langle \wt \varphi_3 \rangle\langle \varphi_3 \rangle \\
\langle \wt \varphi_1 \rangle\langle \varphi_3 \rangle+\omega \langle \wt \varphi_2 \rangle\langle \varphi_1 \rangle+\omega^2 \langle \wt \varphi_3 \rangle\langle \varphi_2 \rangle \\
\langle \wt \varphi_1 \rangle\langle \varphi_2 \rangle+\omega \langle \wt \varphi_2 \rangle\langle \varphi_3 \rangle+\omega^2 \langle \wt \varphi_3 \rangle\langle \varphi_1 \rangle 
\end{pmatrix}=\begin{pmatrix} 0\\0\\0\end{pmatrix} \!,~~~~~ \label{eq:F3}  \\
\nonumber\\ 
\frac{\partial W_{\mathrm{flav}}} {  \partial D_{\bf{\ol 3}}}&\!=\!&M_2\begin{pmatrix} \langle \varphi_1 \rangle\\\langle \varphi_2 \rangle\\\langle \varphi_3 \rangle\end{pmatrix}+
\begin{pmatrix} 
\langle \varphi_1 \rangle^2+2\langle \varphi_2 \rangle\langle \varphi_3 \rangle\\\omega(\langle \varphi_3 \rangle^2+2\langle \varphi_1 \rangle\langle \varphi_2 \rangle)\\\omega^2(\langle \varphi_2 \rangle^2+2\langle \varphi_3 \rangle\langle \varphi_1 \rangle)
\end{pmatrix}\nonumber\\
&&+\begin{pmatrix} 
\langle \wt \varphi_1 \rangle^2-\langle \wt \varphi_2 \rangle\langle \wt \varphi_3 \rangle\\\langle \wt \varphi_2 \rangle^2-\langle \wt \varphi_3 \rangle\langle \wt \varphi_1 \rangle\\\langle \wt \varphi_3 \rangle^2-\langle \wt \varphi_1 \rangle\langle \wt \varphi_2 \rangle
\end{pmatrix}+
\begin{pmatrix} 
\langle \wt \varphi_1 \rangle\langle \varphi_1 \rangle+\omega^2 \langle \wt \varphi_2 \rangle\langle \varphi_2 \rangle+\omega \langle \wt \varphi_3 \rangle\langle \varphi_3 \rangle \\
\langle \wt \varphi_3 \rangle\langle \varphi_1 \rangle+\omega^2 \langle \wt \varphi_1 \rangle\langle \varphi_2 \rangle+\omega \langle \wt \varphi_2 \rangle\langle \varphi_3 \rangle \\
\langle \wt \varphi_2 \rangle\langle \varphi_1 \rangle+\omega^2 \langle \wt \varphi_3 \rangle\langle \varphi_2 \rangle+\omega \langle \wt \varphi_1 \rangle\langle \varphi_3 \rangle 
\end{pmatrix}=\begin{pmatrix} 0\\0\\0\end{pmatrix} \!. ~~~~~\label{eq:F4}
\end{eqnarray}}\normalsize
One can easily see that the desired alignments given in \ref{eq:alignments}
are inconsistent with \ref{eq:F1} and \ref{eq:F2}. \ref{eq:F3} also does not
admit \ref{eq:alignments} as a solution. Turning to \ref{eq:F4}, one can
substitute $\langle \wt \varphi_1 \rangle=v_{\wt\varphi}$, 
$\langle \wt \varphi_2 \rangle=\langle \wt \varphi_3 \rangle=0$ and $\langle \varphi_1 \rangle=\langle \varphi_2 \rangle=\langle \varphi_3 \rangle=\frac{v_{\varphi}}{\sqrt{3}}$ 
to obtain 
\begin{eqnarray}
\mbox{$\frac{1}{\sqrt{3}}$}M_2 v_\varphi+v_\varphi^2+ v_{\wt \varphi}^2+\mbox{$\frac{1}{\sqrt{3}}$}v_{\wt \varphi}  v_\varphi&=&0 \ ,\\
 \mbox{$\frac{1}{\sqrt{3}}$}M_2v_\varphi+\omega v_{\varphi}^2+\mbox{$\frac{1}{\sqrt{3}}$}\omega^2 v_{\wt \varphi} v_\varphi&=&0 \ ,\\
 \mbox{$\frac{1}{\sqrt{3}}$}M_2v_\varphi+\omega^2{v_\varphi}^2+\mbox{$\frac{1}{\sqrt{3}}$}\omega v_{\wt \varphi} v_\varphi&=&0 \ .
\end{eqnarray}
Since $v_\varphi=0$ is not of physical interest, we can cancel off $v_\varphi$ as
common factor in the second and third equation,
which can then be solved easily to obtain $v_\varphi=\frac{M_2}{\sqrt{3}}$ and 
$v_{\wt \varphi}= M_2$. Substituting this into the first equation, yields
$2M_{2}^2=0$, which in turn would require $M_2=0$. Hence, the desired non-trivial
alignment is also incompatible with \ref{eq:F4}.

\medskip

So far we have assumed that {\it all} terms in \ref{eq:F3} and  \ref{eq:F4} are
present. However, in extra dimensional models it is possible to envisage a setup in
which the two flavons $\wt \varphi$ and $\varphi$ are localized on different
branes \cite{Altarelli:2005yp,Callen:2012kd}. Then, a given driving field
could, in principle, couple exclusively to the $\wt \varphi$ flavon, while another
driving field could couple exclusively to $\varphi$. If both driving fields
transform in the ${\bf 3}$ of $T_7$, it is straightforward (though tedious for
the alignment of $\wt \varphi$) to verify that the
two flavons feature the desired alignments up to $T_7$ transformed solutions.

\medskip

We conclude that the $F$-term conditions of~\ref{eq:F1}-\ref{eq:F4} cannot
give rise to the alignments of \ref{eq:alignments} without a mechanism that
effectively sets some of the coupling constants to zero.

\subsection{Alignment through a hidden flavon sector}

One of the attractive features of the $T_7$ model presented in \ref{sec:TBMLO}
is that it does not require the introduction of an extra $\mathbb{Z}_N$ shaping
symmetry to arrive at tribimaximal mixing at leading order. Unfortunately, as
we have found in \ref{sec:vevgd}, the flavons cannot be aligned without
setting some coupling constants of the flavon superpotential to zero.
If we do not want to resort to extra dimensions, we are forced to extend the
minimal $T_7$ symmetry by introducing a new symmetry. However, 
we will show in this section how this can be achieved {\it without} assigning
non-trivial charges under this new symmetry to the flavons $\wt \varphi$ and
$\varphi$, nor to the lepton and Higgs fields. 

\medskip

To this end, it is necessary to introduce new flavons which do not couple to
the leptons. In that sense, one might call them ``hidden flavons'',
constituting a sequestered sector of the model. This separation is achieved by
assuming these flavons (together with some new driving fields) to be the only
fields which transform non-trivially under a hidden
$\mathbb{Z}_N^{\text{hid}}$ symmetry. The flavons $\wt \varphi$ and
$\varphi$  of the lepton (or visible) sector can then be aligned with respect
to those hidden flavons via orthogonality conditions as we discuss now.

\medskip

In the hidden sector, we introduce four flavons $\chi$, $\xi'$, $\psi$, $\wt \zeta$ which are
aligned by virtue of four driving fields $\wt D_{\chi}$, $D_{\psi}$,
$O_{\chi\wt \zeta}$, $O_{\psi\wt \zeta}$. Their transformation properties are
given in \ref{tab:charge_hidden}, where $x,y,z$ are positive integers smaller
than $N$.
\begin{table}[t]
\begin{center}
\begin{tabular}{|c||c|c|c|c||c|c|c|c|}\hline
$\phantom{\Big|}$Field$\phantom{\Big|}$ & 
$\chi$ & $\xi'$ & $\psi$ & $\wt \zeta$  &
$\wt D_{\chi}$ & $D_{\psi}$ & $O_{\chi\wt \zeta}$ & $O_{\psi\wt \zeta}$\\\hline
$\phantom{\Big|}$$T_7$$\phantom{\Big|}$ & 
${\bf 3}$ & ${\bf 1'}$ & ${\bf 3}$ & ${\bf \ol 3}$ & 
${\bf \ol 3}$ & ${\bf 3}$ & ${\bf  1}$& ${\bf 1}$  \\\hline
$\phantom{\Big|}\mathbb Z_N^\mathrm{hid}\phantom{\Big|}$ &
$x$&$x$&$y$&$z$&
$-2x$ & $-2y$ & $-x-z$ & $-y-z$ \\\hline
$\phantom{\Big|}\U{1}_R\phantom{\Big|}$ &
$0$&$0$&$0$&$0$&
$2$ & $2$ & $2$ & $2$ \\\hline
\end{tabular}
\setcapindent{0em}
\caption{The charge assignments of the hidden sector. $x,y,z$ are
positive integers smaller than $N$. The smallest possible value of $N$ is 6 with
$(x,y,z)=(2, 1, 5)$ or equivalently $(4, 5, 1)$, see discussion at the end of this section.}
\label{tab:charge_hidden}
\end{center}
\end{table}
The minimal $\mathbb Z_N^{\mathrm{hid}}$ symmetry and the corresponding values
for $x,y$ and $z$ will be determined at the end of this section. The resulting
terms of the renormalizable hidden flavon superpotential read
\bea
W_{\mathrm{flav}}^{\mathrm{hid}} &= &
\wt D_{\chi} \left(\chi\xi' + \chi \chi   \right) 
\,+\, D_{\psi}  \psi \psi 
\,+\, O_{\chi\wt \zeta}   \chi\wt \zeta  
\,+\, O_{\psi \wt \zeta}   \psi\wt \zeta  \ .
\label{eq:hiddenpot1}
\eea

Setting the $F$-terms of the driving fields to zero gives rise to the conditions
\bea
\frac{\partial  W_{\mathrm{flav}}^{\mathrm{hid}} }{  \partial \wt D_{\chi}} & = & 
\begin{pmatrix} 
\vevof{\chi_2}^{} \\ 
\vevof{\chi_3}^{} \\ 
\vevof{\chi_1}^{}
\end{pmatrix}\vevof{ \xi'}
+ 
\begin{pmatrix} 
\vevof{\chi_1}^{2}+ 2 \vevof{\chi_2}^{} \vevof{\chi_3}^{}\\ 
\omega(\vevof{\chi_3}^{2}+ 2 \vevof{\chi_1}^{}\vevof{\chi_2}^{})\\ 
\omega^2(\vevof{\chi_2}^{2}+ 2 \vevof{\chi_3}^{} \vevof{\chi_1}^{})
\end{pmatrix}
=
\begin{pmatrix} 
0 \\ 
0 \\ 
0
\end{pmatrix}\ \label{eq:hiddenchi},\\
\frac{\partial  W_{\mathrm{flav}}^{\mathrm{hid}}} {  \partial D_{\psi}} & = & 
\begin{pmatrix} 
\vevof{\psi_1}^{2} -\vevof{\psi_2}^{} \vevof{\psi_3}^{} \\ 
\vevof{\psi_2}^{2} -\vevof{\psi_3}^{} \vevof{\psi_1}^{} \\ 
\vevof{\psi_3}^{2} -\vevof{\psi_1}^{} \vevof{\psi_2}^{}
\end{pmatrix}
=
\begin{pmatrix} 
0 \\ 
0 \\ 
0
\end{pmatrix} \ \label{eq:hiddenpsi},\\
\frac{\partial  W_{\mathrm{flav}}^{\mathrm{hid}} }{  \partial O_{\chi \wt \zeta}} & = & 
\vevof{\chi_1}^{} \vevof{\wt \zeta_1}^{} + \vevof{\chi_2}^{} \vevof{\wt \zeta_2}^{} + \vevof{\chi_3}^{} \vevof{\wt \zeta_3}^{}
= 0 \ \label{eq:hiddenzeta1},\\
\frac{\partial  W_{\mathrm{flav}}^{\mathrm{hid}} }{  \partial O_{\psi \wt \zeta}} & = & 
\vevof{\psi_1}^{} \vevof{\wt \zeta_1}^{} + \vevof{\psi_2}^{} \vevof{\wt \zeta_2}^{} + \vevof{\psi_3}^{} \vevof{\wt \zeta_3}^{}
= 0 \ .
\label{eq:hiddenzeta2}
\eea
These conditions can be solved exactly to give a total of 21 sets of
solutions for the hidden flavon alignments. A brief description of how these can
be derived is outlined below. 

\medskip

First, we consider~\ref{eq:hiddenpsi}. Without loss of generality we can
distinguish two cases: $\vevof{\psi_1}=0$ and $\vevof{\psi_1}\neq 0$. 
In the former case one quickly sees that it requires vanishing values for all
the components of $\vevof{\psi}$, thus giving only the trivial vacuum.
For $\vevof{\psi_1}\neq 0$, we obtain three solutions,
$\vevof{\psi}\propto(1,1,1)^T$,
$\vevof{\psi}\propto(1,\omega^2,\omega)^T$ and
$\vevof{\psi}\propto(1,\omega,\omega^2)^T$, which are related to each other by
the $T_7$ transformation $d$. We remark that a $c$~transformation applied to
these three solutions does not modify the alignments but only multiplies them
with an irrelevant overall phase.

\medskip

Next, we turn to~\ref{eq:hiddenchi}. Again there are two cases to distinguish, 
$\vevof{\chi_1}=  0$ and $\vevof{\chi_1}\neq 0$. For the former
case the solution is easily found to be $\vevof{\chi} \propto (0,0,1)^T$. With 
$\vevof{\chi_1}\neq 0$ on the other hand, it is possible to show that there
exist six different solutions, all of which are related to the previous
solution by the $T_7$ symmetry transformation~$c$: $\vevof{\chi} \propto c^k
(0,0,1)^T$, where $k=0,1,...,6$. One can check that applying a
$d$~transformation to any of these seven solutions does not yield new
alignments.

\medskip

Combining our results so far, we have 21 different pairs of alignments 
$\vevof{\chi}$ and  $\vevof{\psi}$. Being related by the 21 different $T_7$ symmetry
transformations, they are physically equivalent and one can choose one pair
without loss of generality. Having fixed the alignments of $\vevof{\chi}$ and
$\vevof{\psi}$, \ref{eq:hiddenzeta1} and \ref{eq:hiddenzeta2} determine the
alignment of $\vevof{\wt\zeta}$ uniquely to be perpendicular to both
$\vevof{\chi}$ and $\vevof{\psi}$. This shows that in total there exist 21
sets of solutions for the hidden flavon alignments. For convenience we choose
the following simple set
\be
\langle \chi \rangle 
\propto \begin{pmatrix} 0\\0\\1\end{pmatrix} , \qquad
\langle \psi \rangle
\propto \begin{pmatrix} 1\\1\\1\end{pmatrix} , \qquad
\langle \wt\zeta \rangle
\propto \begin{pmatrix} 1\\-1\\0\end{pmatrix}  .
\ee

Let us now turn to the discussion of aligning the original flavon fields $\wt
\varphi$ and $\varphi$ which enter in the lepton sector. 
Their alignment is dictated by new driving fields which couple the pre-aligned
hidden flavons to $\wt \varphi$ and $\varphi$. The transformation 
properties of the extra driving fields $O_{\chi \wt\varphi}$, $O'_{\chi \wt\varphi}$, 
$O_{\wt \zeta \varphi}$ and $O'_{\wt \zeta \varphi}$ are listed
in~\ref{tab:hiddendriv}. 
\begin{table}[t]
\begin{center}
\begin{tabular}{|c||c|c|c|c|}\hline
$\phantom{\Big|}$Field$\phantom{\Big|}$ & 
$O_{\chi \wt\varphi}$ & $O'_{\chi \wt\varphi}$ & 
$O_{\wt \zeta \varphi}$ & $O'_{\wt \zeta \varphi}$\\\hline
$\phantom{\Big|}$$T_7$$\phantom{\Big|}$ & 
${\bf 1}$ & ${\bf 1'}$ & ${\bf 1}$ & ${\bf 1'}$  \\\hline
$\phantom{\Big|}\mathbb Z_N^\mathrm{hid}\phantom{\Big|}$ &
$-x$&$-x$&$-z$&$-z$\\\hline
$\phantom{\Big|}\U{1}_R\phantom{\Big|}$ &
$2$ & $2$ & $2$ & $2$ \\\hline
\end{tabular}
\end{center}
\setcapindent{0em}
\caption{Charge assignments of the driving fields coupling the hidden flavons
  $\chi$, $\wt \zeta$
  to the flavons $\wt\varphi$, $\varphi$.}
\label{tab:hiddendriv}
\end{table}
With these assignments, the resulting renormalizable flavon superpotential
consisting of all $T_7$ and $\mathbb Z_N^\mathrm{hid}$ invariant terms reads
\bea
W_{\mathrm{flav}}^{\prime} &= &
O_{\chi \wt\varphi} \chi \wt\varphi 
\,+\, O'_{\chi \wt\varphi} \chi \wt\varphi 
\,+\,O_{\wt \zeta \varphi}\wt \zeta \varphi
\,+\,O'_{\wt \zeta \varphi}\wt \zeta \varphi \  .
\label{eq:hiddenpot2}
\eea
The derived $F$-term conditions take the form
\bea
\frac{\partial W_{\mathrm{flav}}^{\prime} }{  \partial O_{\chi \wt\varphi}} &= &
\langle \chi_1 \rangle^{} \langle \wt\varphi_1  \rangle^{} + \langle \chi_2 \rangle^{} \langle \wt\varphi_2  \rangle^{} + \langle \chi_3 \rangle^{}
\langle \wt\varphi_3  \rangle^{} = \langle \chi_3 \rangle^{}
\langle \wt\varphi_3  \rangle^{} = 0  \ \label{FF1}, \\
\frac{\partial W_{\mathrm{flav}}^{\prime} }{  \partial O'_{\chi \wt\varphi}} &= &
\langle \chi_1 \rangle^{} \langle \wt\varphi_3  \rangle^{} + \langle \chi_2 \rangle^{} \langle \wt\varphi_1  \rangle^{} + \langle \chi_3 \rangle^{}
\langle \wt\varphi_2  \rangle^{} =   \langle \chi_3 \rangle^{}
\langle \wt\varphi_2  \rangle^{} = 0 \ \label{FF2}, \\
\frac{\partial W_{\mathrm{flav}}^{\prime}}{  \partial O_{\wt \zeta \varphi}}&= &
\langle \wt \zeta_1  \rangle^{} \langle \varphi_1  \rangle^{} + \langle \wt \zeta_2  \rangle^{} \langle \varphi_2  \rangle^{} + \langle \wt \zeta_3  \rangle^{}
\langle \varphi_3  \rangle^{} = 
\langle \wt \zeta_1  \rangle^{} ( \langle \varphi_1  \rangle^{} - \langle \varphi_2  \rangle^{} ) = 0  \ \label{FF3} , \\
\frac{\partial W_{\mathrm{flav}}^{\prime} }{  \partial O'_{\wt \zeta \varphi}}&= &
\langle \wt \zeta_3  \rangle^{} \langle \varphi_1  \rangle^{} + \langle \wt \zeta_1  \rangle^{} \langle \varphi_2  \rangle^{} + \langle \wt \zeta_2  \rangle^{}
\langle \varphi_3  \rangle^{} =  \langle \wt \zeta_1  \rangle^{} ( \langle \varphi_2  \rangle^{} - \langle \varphi_3  \rangle^{} ) = 0 \ \label{FF4} ,
\eea
yielding the desired alignments, cf. \ref{eq:alignments},
\be
\langle \wt \varphi \rangle 
\propto \begin{pmatrix} 1\\0\\0\end{pmatrix} , \qquad
\langle \varphi \rangle
\propto \begin{pmatrix} 1\\1\\1\end{pmatrix} . 
\ee

Finally, we need to choose an appropriate $\mathbb{Z}_N^{\text{hid}}$,
i.e. suitable values for $N$, $x$, $y$ and $z$, so that no additional
renormalizable terms, other than the ones discussed in this section, occur in the
flavon superpotential of \ref{eq:hiddenpot1} and \ref{eq:hiddenpot2}.
Through a systematic search, we find that the smallest
$\mathbb{Z}_N^{\text{hid}}$ that we can have is a $\mathbb{Z}_6^{\text{hid}}$
[although \ref{eq:hiddenpot1} alone can be obtained using a
$\mathbb{Z}_5^{\text{hid}}$].  For a $\mathbb{Z}_6^{\text{hid}}$, we can
have $(x,y,z)$ as $(2,1,5)$ or $(4,5,1)$. Note that these two choices are
equivalent solutions since the elements of the group corresponding to one set
are the negatives of the elements corresponding to the other.

\newpage

\section{Conclusions}
\label{sec:concl}

In the present work, we discussed a minimal model of neutrino flavor. In this context, the term ``minimal'' refers to the order of our flavor group $T_7$, and to the number of flavon fields. It is interesting to note that the $\mathbb{Z}_2\times\mathbb{Z}_2$ Klein symmetry is not a subgroup of $T_7$, but arises completely accidentally. As such, it is -- to the best of our knowledge -- the first indirect model in which the flavon fields appear linearly (and not quadratically) in the leading order structure of the model. Another striking feature of our model is the absence of any Abelian shaping symmetries like $\mathbb{Z}_N$ or \U{1}. As a consequence, our flavor group $T_7$ with its 21 elements is smaller than any symmetry $A_4\times\mathbb{Z_N}$ that has long been a paradigm for model building before Daya Bay's and Reno's measurements of a large reactor angle $\theta_{13}\simeq9^\circ$. Our model predicts tribimaximal mixing at leading order, and we achieve a sizable $\theta_{13}$ by considering next-to-leading order corrections to the superpotential. It turns out that models in full agreement with experiment can be obtained for some generic values of the flavon \vevs{} and coefficients in the superpotential. More elegantly, however, one can isolate the next-to-leading order contribution proportional to the coefficient $\Cfive{}$. Being of trimaximal type, this correction drives $\theta_{13}$ to sizable values while leaving the solar angle very close to its tribimaximal value. The atmospheric mixing angle is correlated to the size of $\theta_{13}$ and stays within the allowed $3\sigma$ region. In \ref{sec:uv-mo}, we embedded our effective theory into a renormalizable model that naturally suppresses all next-to-leading order contributions except the trimaximal one which is responsible for the large reactor angle. In the last section, we discussed how the flavon fields can be dynamically aligned to yield the required symmetry breaking pattern. The alignment of the flavon fields necessitated the introduction of an Abelian shaping symmetry and so-called driving fields. However, one should note that the $F$-term alignment we considered in \ref{sec:vev-4} is only one possible option and that the details of an elegant and simple mechanism of vacuum stabilization are still lurking in the dark.


\section*{Acknowledgments}

We acknowledge useful discussions with Steve King and Stuart
Raby. C.L.~acknowledges partial support from the EU ITN grants UNILHC
237920 and INVISIBLES 289442 and would like to thank the LPSC Grenoble
for their hospitality. K.P.~is supported by the Shyama Prasad
Mukherjee Fellowship from the Council of Scientific and Industrial
Research (CSIR), India. C.L.~and A.W.~would like to thank the
\emph{Galileo Galilei Institute for Theoretical Physics} in Florence
for their hospitality. We are greatly indebted to the \emph{Centre de
Calcul de l'Institut National de Physique Nucl\'{e}aire et Physique
des Particules} in Lyon for using their resources.


\appendix

\labelformat{section}{Appendix #1} 

\section{Generators and Clebsch-Gordan coefficients of $\bs{T_7}$}
\label{app:A}

In this appendix we list the relevant group theory of $T_7$, which is
sometimes also called the Frobenius group $\mathbb Z_7 \rtimes \mathbb
Z_3$. The group is obtained from two generators $c,d$ obeying the
presentation, see e.g.~\cite{Luhn:2007yr},
\be
<c,d\,|\,c^7=d^3=1 \,,\, d^{-1}cd=c^4 > \ . \label{presentation}
\ee
It has 21 elements and five irreducible representations, namely 
${\bf 1}$, ${\bf 1'}$, ${\bf 1''}$, ${\bf 3}$, and ${\bf \ol 3}$. 
A pair of triplet generators satisfying the presentation
in~\ref{presentation} can be found e.g. 
in~\cite{Luhn:2007yr,Hagedorn:2008bc,King:2009ap}, where the  
order-seven generator $c$ is diagonal. For the purpose of the model in the
present paper, it is however more convenient to work in a basis with a
diagonal order-three generator $d$. Our choice of $T_7$ generators for all five
irreducible representations is listed in~\ref{tab:basis}, where we have
defined 
\be
\omega = e^{\frac{2\pi i}{3}} \ ,\qquad
\eta = e^{\frac{2\pi i}{7}} \ .
\ee

\begin{table}[t]
\begin{center}
$
\begin{array}{ccc}
& d & c \\[1mm] \hline \\[-3mm]
{\bf 1} & 1& 1 \\[3mm]
{\bf 1'} & \omega  & 1 \\[3mm]
{\bf 1''} & \omega^2& 1 \\[3mm]
{\bf 3} & \begin{pmatrix} 1&0&0\\0&\omega^2&0\\ 0&0&\omega  \end{pmatrix} & 
\frac{\eta}{3} \begin{pmatrix} 
1+\eta+\eta^3 &
\omega^2+\omega \eta   + \eta^3&
\omega+\omega^2 \eta+\eta^3 \\
\omega+\omega^2 \eta+\eta^3 &
1+\eta+\eta^3&
\omega^2+\omega \eta   + \eta^3\\
\omega^2+\omega \eta   + \eta^3 & 
\omega+\omega^2 \eta+\eta^3 &
1+\eta+\eta^3 
\end{pmatrix}
 \\\\[3mm]
{\bf \ol 3} & \begin{pmatrix} 1&0&0\\0&\omega&0\\ 0&0&\omega^2  \end{pmatrix} & 
\frac{\eta^6}{3} \begin{pmatrix} 
1+\eta^6+\eta^4 &
\omega+\omega^2 \eta^6   + \eta^4&
\omega^2+\omega \eta^6+\eta^4 \\
\omega^2+\omega \eta^6+\eta^4 &
1+\eta^6+\eta^4&
\omega+\omega^2 \eta^6   + \eta^4\\
\omega+\omega^2 \eta^6   + \eta^4 & 
\omega^2+\omega \eta^6+\eta^4 &
1+\eta^6+\eta^4 
\end{pmatrix}
\end{array}
$
\end{center}
\setcapindent{0em}
\caption{\label{tab:basis}The $T_7$ generators of the five irreducible
  representations in the basis where $d$ is diagonal. 
Here we have defined $\omega = e^{\frac{2\pi i}{3}}$ and 
$\eta = e^{\frac{2\pi i}{7}}$.} 
\end{table}

Although the order-seven generators of the triplet representations look rather
involved, the Clebsch-Gordan coefficients take a relatively simple
form. Omitting the trivial products, i.e. those involving the singlet ${\bf 
  1}$ as well as products of only one-dimensional irreducible representations,
the product rules are reported below. We use the convention that the components of
the first representation of any given product ${\bf a\otimes b}$ are denoted
by $a_i$ while we use $b_i$ for the components of the second
representation. The subscripts $s$ and $a$ stand for ``symmetric'' and
``anti-symmetric'', respectively. Note that we have also included the
normalization factors of the Clebsch-Gordan coefficients.

\be
{\bf 1' \otimes  3}:
\quad  {\bf 3} ~=~ a_1 \begin{pmatrix} 
b_2\\b_3\\b_1 
 \end{pmatrix} \ , \qquad \qquad
{\bf 1'' \otimes  3}:
\quad {\bf 3} ~=~ a_1 \begin{pmatrix} 
b_3\\b_1\\b_2 
 \end{pmatrix} \ ,
\ee

\vspace{3mm}

\be
{\bf 1' \otimes  \overline{3}}:
\quad  {\bf \overline{3}} ~=~ a_1 \begin{pmatrix} 
b_3\\b_1\\b_2 
 \end{pmatrix} \ , \qquad\qquad
{\bf 1'' \otimes  \overline{3}}:
\quad  {\bf \overline{3}} ~=~ a_1 \begin{pmatrix} 
b_2\\b_3\\b_1 
 \end{pmatrix} \ ,
\ee

\bea
{\bf 3 \otimes \ol 3}:
&& {\bf 1} ~=~ \mbox{$\frac{1}{\sqrt{3}}$}\left(a_1b_1+a_2 b_2+a_3b_3\right) \ , \\[2mm]
&&  {\bf 1'} ~=~  \mbox{$\frac{1}{\sqrt{3}}$}\left(a_1b_2+a_2 b_3+a_3b_1\right) \ , \\[2mm]
&&  {\bf 1''} ~=~   \mbox{$\frac{1}{\sqrt{3}}$}\left(a_1b_3+a_2 b_1+a_3b_2\right) \ , \\[1mm]
&&  {\bf 3} ~=~  \mbox{$\frac{1}{\sqrt{3}}$}\begin{pmatrix} 
a_1b_1+\omega^2 a_2 b_2+\omega a_3b_3  \\
a_1b_3+\omega^2 a_2 b_1+\omega a_3b_2 \\
a_1b_2+\omega^2 a_2 b_3+\omega a_3b_1
 \end{pmatrix} \ ,\\[2mm]
&&  {\bf \ol 3} ~=~ \mbox{$\frac{1}{\sqrt{3}}$}\begin{pmatrix} 
a_1b_1+\omega a_2 b_2+\omega^2 a_3b_3  \\
a_3b_1+\omega a_1b_2+\omega^2 a_2 b_3  \\
 a_2 b_1+\omega a_3b_2+\omega^2 a_1b_3
 \end{pmatrix} \ ,
\eea

\bea
~\,{\bf 3 \otimes  3}:
&&  {\bf 3}_s ~=~\mbox{$\frac{1}{\sqrt{3}}$} \begin{pmatrix} 
a_1b_1+ a_2 b_3+ a_3b_2  \\
\omega (a_1b_2+ a_2 b_1+a_3b_3) \\
\omega^2 (a_1b_3+ a_2 b_2+ a_3b_1)
 \end{pmatrix} \ ,\\[2mm]
&&  {\bf \ol 3}_s ~=~\mbox{$\frac{1}{\sqrt{6}}$}\begin{pmatrix} 
2 a_1b_1- a_2 b_3- a_3b_2  \\
2 a_2b_2- a_3 b_1- a_1b_3  \\
2 a_3b_3- a_1 b_2- a_2b_1
 \end{pmatrix} \ ,\\[2mm]
&&  {\bf \ol 3}_a ~=~\mbox{$\frac{1}{\sqrt{2}}$}\begin{pmatrix} 
 a_2 b_3- a_3b_2  \\
 a_3 b_1- a_1b_3  \\
 a_1 b_2- a_2b_1
 \end{pmatrix} \ .
\eea


\bibliography{mybibliography}

\bibliographystyle{utphys}

\end{document}